\documentclass[conference]{IEEEtran}
\usepackage{cite}
\usepackage{amsmath}
\usepackage{enumitem}
\usepackage{amssymb}
\usepackage{amsthm}
\usepackage{graphicx,epstopdf}
\usepackage{multicol}
\usepackage{textcomp}
\usepackage{multirow}
\usepackage{flushend}
\newcommand\T{\rule{0pt}{2.4ex}}

\usepackage{color}
\makeatletter
\newcommand*\bigcdot{\mathpalette\bigcdot@{0.85}}
\newcommand*\bigcdot@[2]{\mathbin{\vcenter{\hbox{\scalebox{#2}{$\m@th#1\bullet$}}}}}
\makeatother
\usepackage{lipsum}
\usepackage{cuted}
\usepackage{tikz}
\usepackage{wrapfig, blindtext}
\usepackage{hyperref}

\title{Power System Dynamic State Estimation Using Extended and Unscented Kalman Filters \vspace{-0.5ex}}

\author{Narayan Bhusal and  Mukesh Gautam\\
Department of Electrical and Biomedical Engineering, 
University of Nevada, Reno, NV 89557, USA\\
Emails: bhusalnarayan62@nevada.unr.edu and mukesh.gautam@nevada.unr.edu\vspace{-3ex}}

\begin{document}
\maketitle
\thispagestyle{empty}
\pagestyle{empty}

\begin{abstract}
Accurate estimation of power system dynamics is very important for the enhancement of power system reliability, resilience, security, and stability of power system. With the increasing integration of inverter-based distributed energy resources, the knowledge of power system dynamics has become more necessary and critical than ever before for proper control and operation of the power system. Although recent advancement of measurement devices and the transmission technologies have reduced the measurement and transmission error significantly, these measurements are still not completely free from the measurement noises. Therefore, the noisy measurements need to be filtered to obtain the accurate power system operating dynamics.  In this work, the power system dynamic states are estimated using extended Kalman filter (EKF) and unscented Kalman filter (UKF).  We have performed case studies on Western Electricity Coordinating Council (WECC)'s $3$-machine $9$-bus system and New England $10$-machine $39$-bus. The results show that the UKF and EKF can accurately estimate the power system dynamics. The comparative performance of EKF and UKF for the tested case is also provided. {\color{red} Other Kalman filtering techniques along with the machine learning based estimator will be updated in this report soon.} \textit{All the sources code including Newton Raphson power flow, admittance matrix calculation, EKF calculation, and UKF calculation are publicly available in Github on } \href{https://github.com/nbhusal/Power_System_Dynamic_State_Estimation}{Power System Dynamic State Estimation.}
\end{abstract}
\IEEEpeerreviewmaketitle

\section{INTRODUCTION}
Conventionally, power system state estimation (PSEE) used to be performed by static approaches based on weighted least square (WLS) method, in which a single set of measurements is used to estimate the system states. The WLS based methods have been widely used because of their simplicity and fast convergence. But the static state estimators cannot predict the future operating points of the system even when the accuracy of the estimation is within acceptable limits under fully observable conditions \cite{rousseaux1990whither}. Moreover, with the increasing penetration of distributed energy resources (DERs), responsive loads, and microgrids, the power system have been subjected to different types of dynamics. For example, the stochastic and intermittent characteristics of DERs increase the possibility of rapid changes in the bus voltages within short time-frame \cite{hassanzadeh2015short}. These changes may lead to the changes in active and reactive power resulting in the change in generator's state variables such as rotor angle and speed. The majority of monitoring and control tools that are currently available at control center are based on static state estimation, which may not be capable to capture such dynamics \cite{kamwa2016synchrophasors}. Therefore, an estimator with very high accuracy that gives can continuously track the dynamic changes in the non-linear power systems is required.

Since power systems are regarded as quasi-static system under normal conditions, slow and steady changes take place in the system which are mainly driven by the system loads. Because of the changes in the system loads, the generations are also adjusted accordingly. In order to capture this dynamics, the state estimation must be performed at short intervals of time. Dynamic state estimators effectively fit this purpose. Dynamic state estimation (DSE) algorithms have the potential to impact the operation of the real time monitoring and control of power systems \cite{shivakumar2008review}. 

Different methods have been applied in the literature for the implementation of dynamic state estimation (DSE) of power system problems. A robust Kalman filter has been developed for DSE of power system using model transformation in \cite{wang2020fast}, in which the proposed model transformation strategy is achieved by taking the measured generator active power as the input variable and the derived frequency and the rate of change of frequency measurements from the phasor measurement units (PMUs) as the output variables of the dynamical generator model. In \cite{zhao2017power}, an unscented Kalman filter (UKF)-based DSE has been proposed by integrating measurement correlations, in which the correlations between voltage phasors and calculated real/reactive power as well as the correlations between real and reactive power at the generator terminal buses have been analyzed. In \cite{tebianian2015dynamic} extended kalman fillter has been used to estimate the power system states. In \cite{zhao2016robust}, a robust iterative extended Kalman filter (EKF) based on the generalized maximum likelihood approach has been proposed for estimating power system state dynamics when subjected to disturbances, which can track the system transients in faster and more reliable way than the conventional EKF and UKF. A robust generalized maximum-likelihood-based UKF has been presented in \cite{zhao2017robust}, where the statistical linearization approach is used to derive a compact batch-mode regression form by processing the predicted state vector and the received measurements simultaneously. In \cite{goleijani2018multi}, a multi-agent based modeling for DSE of power system has been proposed that is able to take advantages of hybrid measurement data.

In this work, we have estimated the power system dynamic states using extended Kalman filter (EKF) and unscented Kalman filter (UKF) technique.  We have performed case studies on western electricity coordinating council's $3$ machine $9$ bus system and New England $10$ machine $39$ bus. The result show that the UKF and EKF can accurately estimate the power system dynamics. The comparative performance of EKF and UKF for the tested case is also provided. All the source code written in MATLAB programming environment are publicly available to help the beginner understand the power system dynamic state estimation. 

The rest of the paper is organized as follows. Section \ref{sec:EKF} briefly describes the preliminaries of the EKF. Section \ref{sec:UKF} describes about the UKF. Section \ref{sec:power_system_dynamics} provides brief description of power system dynamics. Section \ref{sec:simulation_verification} provides the case study to show the performance of EKF and UKF for power system dynamics state estimation. Finally concluding remarks are provided in \ref{sec:conclusion}.

\section{Extended Kalman Filter for Dynamic State Estimation}\label{sec:EKF}
Since the basic Kalman filter is limited to a linear assumption, it is generally extended when there is nonlinearity associated with either the process model or the measurement model or both. The extended Kalman filter (EKF) is the nonlinear version of Kalman filter which linearizes about an estimate of the current mean and error covariance. 

The main steps of the extended Kalman filter are summarized as follows \cite{simon2006optimal}.
\begin{enumerate}
    \item The discrete time system equations of a non-linear system can be presented as 
    \begin{equation}
        \begin{aligned}
            x_{k+1}=&~f_k(x_k,u_k,w_k)\\
               y_k=&~h_k(x_k,v_k)\\
           w_k \sim &~(0,Q_k)\\
            v_k \sim &~(0,R_k)
        \end{aligned}
        \label{equ:non-linear_discrete_system}
    \end{equation}
    \item The initial state of EKF is initialized by taking the expectation of the initial state of the system and the initial state covariance matrix is initialized by taking the second moment of the system state about initial estimate. Mathematically, it can be expressed as follows.
       \begin{equation}
   \begin{aligned}
      & \hat{x}_0^{+}=E(x_0) \\
      & P_0^{+}=E[(x_0-\hat{x}_0^{+}) (x_0-\hat{x}_0^{+})^T]
   \end{aligned}
   \label{equ:non_linear_initialize}
   \end{equation}
   \item For each time step $k$, the prediction of state and error covariance matrix is done as follows.
   \begin{enumerate}
       \item Partial derivative matrices of the current state estimate $\hat{x}^+_{k-1}$ are computed as follows.
       \begin{equation}
           \begin{aligned}
                  F_k=\left.\frac{\partial f_k}{\partial x} \right|_{\hat{x}^+_{k-1}}\\
                   L_k=\left.\frac{\partial f_k}{\partial w} \right|_{\hat{x}^+_{k-1}}
           \end{aligned}
       \end{equation}
       \item The time update of state estimate and estimation-error covariance matrix is performed using:
       \begin{equation}
           \begin{aligned}
                  P_k^- = F_k P^+{k-1} F^T_k + L_k Q_k L_k\\
                  \hat{x}^-_k = f_k(\hat{x}^+_{k-1},u_{k-1},0)
           \end{aligned}
       \end{equation}
   \end{enumerate}
   \item For each time step $k$, the correction of state and error covariance matrix is done as follows.
   \begin{enumerate}
       \item Partial derivative matrices for correction are computed as:
       \begin{equation}
           \begin{aligned}
                H_k=\left.\frac{\partial h_k}{\partial x} \right|_{\hat{x}^-_{k}}\\
                   V_k=\left.\frac{\partial h_k}{\partial v} \right|_{\hat{x}^-_{k}}  
           \end{aligned}
       \end{equation}
       \item The measurement update of the state estimate and estimation error covariance is performed as follows:
       \begin{equation}
           \begin{aligned}
                 & K_k=P^-_k H^T_k (H_k P_k^- H_k^T+V_k R_k V_k^T)^{-1}\\
                  &\hat{x}^+_k =\hat{x}^-_k + K_k[z_k - h_k(\hat{x}^-_k,0)]\\
                 & P^+_k = (I-K_k H_k)P^-_k
           \end{aligned}
       \end{equation}
       
   \end{enumerate}
\end{enumerate}
\section{Unscented Kalman Filter for Dynamic State Estimation}\label{sec:UKF}
When the process model and measurement model are highly nonlinear, the EKF may give poor performance \cite{julier1997new}. This is because of the propagation of the error covariance through linearization of the underlying nonlinear model. In such cases, the unscented Kalman filter (UKF) can be used, which uses a deterministic sampling technique known as the unscented transformation (UT) to generate a minimum set of sample points (referred to as sigma points) around the mean \cite{julier1997new}. These sigma points are then transformed through the nonlinear functions, from which the estimates of new mean and error-covariance are computed. In some of the applications, UKF is applied to reduce the computational cost of the estimation, as there is no requirement to calculate Jacobians.

The main steps of the uncented kalman filter are summarized as follows \cite{simon2006optimal}.
\begin{enumerate}
    \item Let us consider we have the n-state discrete-time system as in \eqref{equ:non-linear_discrete_system}.
   \item We initialize UKF similar to that of EKF using \eqref{equ:non_linear_initialize}. 
   \item To time update the states from one measurement time to another following steps are performed.
   \begin{enumerate}
       \item Choose sigma points to propagate from $k-1$ to $k$ time step using recent best guess of $P$ and $\hat{x}$ as follows. 
       \begin{equation}
           \begin{aligned}
                  \hat{x}_{k-1}^{(i)}= ~&\hat{x}_{k-1}^{+}+\Tilde{x}^{(i)}~~~~ i=1, \cdots, 2n \\
                  \Tilde{x}^{(i)}=&\Bigg(\sqrt{nP_{k-1}^{+}} \Bigg)_i^{T}~~~~ i=1, \cdots, n \\
                  \Tilde{x}^{(i+n)}=&-\Bigg(\sqrt{nP_{k-1}^{+}} \Bigg)_i^{T}~~~~ i=1, \cdots, n
           \end{aligned}
       \end{equation}
   \item Using appropriate changes on the nonlinear function $f(.)$ to transform the sigma points into $\hat{x}_{k}^{(i)}$. 
   \begin{equation}
       \hat{x}_k^{(i)}=f(x_{k-1}^{(i)}, u_k, t_k)
   \end{equation}
   \item Obtain the \textit{priori} state estimate by combining $\hat{x}_k^{(i)}$ vector as follows.
   \begin{equation}
       \hat{x}_k^{-}=\frac{1}{2n}\sum_{i=1}^{2n}\hat{x}_k^{(i)}
   \end{equation}
   \item using the \textit{priori} and $\hat{x}_k^{(i)}$ compute the error covariance matrix. Note that we need to include process noise vector $Q$ to account the process noise.
   \begin{equation}
       P_k^{-}=\frac{1}{2n}\sum_{i=1}^{2n}\Big(\hat{x}_k^{(i)} -\hat{x}_k^{-} \Big) (\hat{x}_k^{(i)} -\hat{x}_k^{-} \Big)^T+Q_{k-1}
   \end{equation}
   \end{enumerate}
   \item Now using the time update equation, we perform the measurement update using following steps. 
   \begin{enumerate}
       \item Determine new sigma point $\hat{x}_k^{(i)}$ by including latest changes in covariance matrix and the estimated state. 
       \begin{equation}
           \begin{aligned}
                  \hat{x}_{k-1}^{(i)}=~&\hat{x}_{k}^{-}+\Tilde{x}^{(i)}~~~~ i=1, \cdots, 2n \\
                  \Tilde{x}^{(i)}=&\Bigg(\sqrt{nP_{k}^{-}} \Bigg)_i^{T}~~~~ i=1, \cdots, n \\
                  \Tilde{x}^{(i+n)}=&-\Bigg(\sqrt{nP_{k}^{-}} \Bigg)_i^{T}~~~~ i=1, \cdots, n
           \end{aligned}
       \end{equation}
       \item Using the latest sigma points on measurement function $h(.)$, determine the predicted measurements, $\hat{z}_k^{(i)}$, as follows. 
       \begin{equation}
           \hat{z}_k^{(i)}=h(\hat{x}_k^{(i)}, t_k)
       \end{equation}
       \item Combine the predicted measurements $\hat{z}_k^{(i)}$ to compute the predicted measurement at time $k$ as follows. 
       \begin{equation}
           \hat{z}_k=\frac{1}{2n}\sum_{i=1}^{2n}\hat{z}_k^{(i)}
       \end{equation}
       \item Compute the covariance of predicted measurements. In this case we need to include $R_k$ to consider the measurement noise. 
       \begin{equation}
           P_z=\frac{1}{2n}\sum_{i=1}^{2n}\Big(\hat{z}_k^{(i)} -\hat{z}_k \Big) (\hat{z}_k^{(i)} -\hat{z}_k \Big)^T+R_{k}
       \end{equation}
       \item Compute cross covariance 
       \begin{equation}
           P_{xz}=\frac{1}{2n}\sum_{i=1}^{2n}\Big(\hat{x}_k^{(i)} -\hat{x}_k \Big) (\hat{z}_k^{(i)} -\hat{z}_k \Big)^T
       \end{equation}
       \item Correct the state as follows. 
       \begin{equation}
           \begin{aligned}
                  K_k=&~P_{xz}P_z^{-1}\\
                  \hat{x}_k^{+}=&~\hat{x}_k^{-}+K_k\left(y_k-\hat{y}_k \right) \\
                  P_k^{+}=&~P_k^{-}-K_kP_zK_k^T
           \end{aligned}
       \end{equation}
   \end{enumerate}
\end{enumerate}

\section{Power System Dynamic Model}\label{sec:power_system_dynamics}
In this section we provide the dynamics of the power system. These dynamics are used  to estimate the dynamic state of the power system. The classical generator model can be expressed as follows.
\begin{equation}
    \begin{aligned}
           &\Dot{\delta}_i=\omega_i-\omega_0  \\
        &\Dot{\omega}_i=\frac{\omega_0}{2H_i} \Big(P_{mi}-P_{Gi}-D\left (\omega_i-\omega_0\right)\Big) 
    \end{aligned}
    \label{equ:power_dynamic}
\end{equation}
where $\delta_i$ is the rotor angle of generator $i$; $\omega_i$ is the angular speed of the generator; $\omega_0$ is the synchronous (rated) speed of the generator; $P_{mi}$ is the mechanical power of generator $i$; $P_{Gi}$ is the electrical power output of generator $i$; $D$ is the damping coefficient; and $H$ is the generator inertia constant. The electrical power output of the generator can be expressed as 
\begin{equation} \label{equ:electrical_real_power}
    P_{Gi}= E_i\sum_{j=1}^n Y_{ij} E_j  \cos{(\delta_{i}-\delta_{j}-\theta_{ij})}
\end{equation}  
where $Y$ is admittance matrix of a reduced network that only have internal generator buses and $E$ is generator internal voltage and $\theta$ is the angles of $Y$. $Y$ can be calculated as follows. 
\begin{equation}
    Y=Y_{22}-Y_{21}\times Y_{11}^{-1}\times Y_{12}
\end{equation}
where $Y_{11}$  is the admittance matrix between the loads; $Y_{12}=Y_{21}^T$ is the admittance matrix between the load and the generator ; $Y_{22}$ is the admittance matrix between the machines. Detail procedure to calculate these matrix is provided in \cite{AndersonFaud77}. 

Equation \eqref{equ:power_dynamic} is  equivalent to $f(.)$ of \eqref{equ:non-linear_discrete_system}, it can be written in discrete form as follows. 
\begin{equation}
    \begin{aligned}
    f(x_{k}, u_k)=
    \begin{cases}
    \delta_{i, k}=\delta_{i, k-1} +{ \omega }_{0}\times ({ \omega }_{i, k-1} -1)+w_{i, \delta } \\
    \omega_{i, k}={ \omega }_{i, k-1} +\Delta t \times ({P}_{mi} -{P}_{Gi,k-1} \\~~~~- { D}( \omega_{i, k-1}-1))/{M}+{w}_{i, \omega } 
    \end{cases}
    \end{aligned}
\end{equation}
where $w_{i, \delta}$ and  $w_{i, \omega}$, respectively, are process noise associated with state $\delta$ and $\omega$ and $\Delta t$ is simulation step size. Therefore, the states to be predicted is machine angle $\delta$ and the machine angular speed $\omega$. 

The measurement nonlinear $h (.)$ in \eqref{equ:non_linear_initialize} for power system can be derived as follows. This is also called as measurement model of the power system dynamic state estimation. For power system dynamic state estimation, generally electrical real and reactive  power obtained from machine and the voltage magnitude and phase angle measurements from each bus are taken as measurements. The expression for electrical real power obtained from the machine is already provided in \eqref{equ:electrical_real_power}. The expression for electrical reactive power output obtained from the machine can be expressed as follows.

\begin{equation} \label{equ:electrical_reactive_power}
    Q_{Gi}=E_i\sum_{j=1}^n Y_{ij} E_j  \sin{(\delta_{i}-\delta_{j}-\theta_{ij})}
\end{equation} 

The voltage magnitude and phase angle measurements can be derived as follows. Let $Y_{exp}$ is the expended system matrix, which can be expressed as \cite{AndersonFaud77}. 
\begin{equation} \label{equ:exp_admittance}
Y_{exp} V_{exp}=
    \begin{pmatrix}
 Y_{11} & Y_{12}\\
 Y_{21} & Y_{22}
\end{pmatrix} 
    \begin{pmatrix}
 V\angle \theta \\
 E\angle \delta 
\end{pmatrix} 
=    \begin{pmatrix}
 0 \\
 I_{G}\angle \delta 
\end{pmatrix}
\end{equation}

where; $V_{exp}$ is expanded voltage vector which includes machine internal voltage $E$ and bus voltage vector $V$. $I_{G}$ is the current injected by the machines. As the load do not inject any current, the upper part of the current injection vector is zero. Using \eqref{equ:exp_admittance}, we can drive the relationship between $V$ and $E$ as follows. 

\begin{equation}
    V\angle \theta = (-Y_{11})^{-1} Y_{22} E\angle \delta = R_V E\angle \delta 
\end{equation}
 where $R_v$ denotes the voltage reconstruction matrix.
 
Therefore, dynamic state estimation and measurement model can be written as follows.  

\begin{gather}
X=\left [\delta^T~~~ \omega^T \right ] \\    
Z=\left [P_{Gi}^T~~~ Q_{Gi}^T~~~ V^T ~~~\theta^T \right ]
\end{gather}

\section{Simulation Verification}\label{sec:simulation_verification}
This section provides the case studies to validate the capability of the extended Kalman filter and the unscented Kalman filter to estimate the power system dynamic states. Case studies results are provided for WECC $3$ machine 9-bus system and New England $10$ machine 39-bus system. 
\subsection{Case WECC 9-bus System}
\begin{figure}
    \includegraphics[scale=0.3]{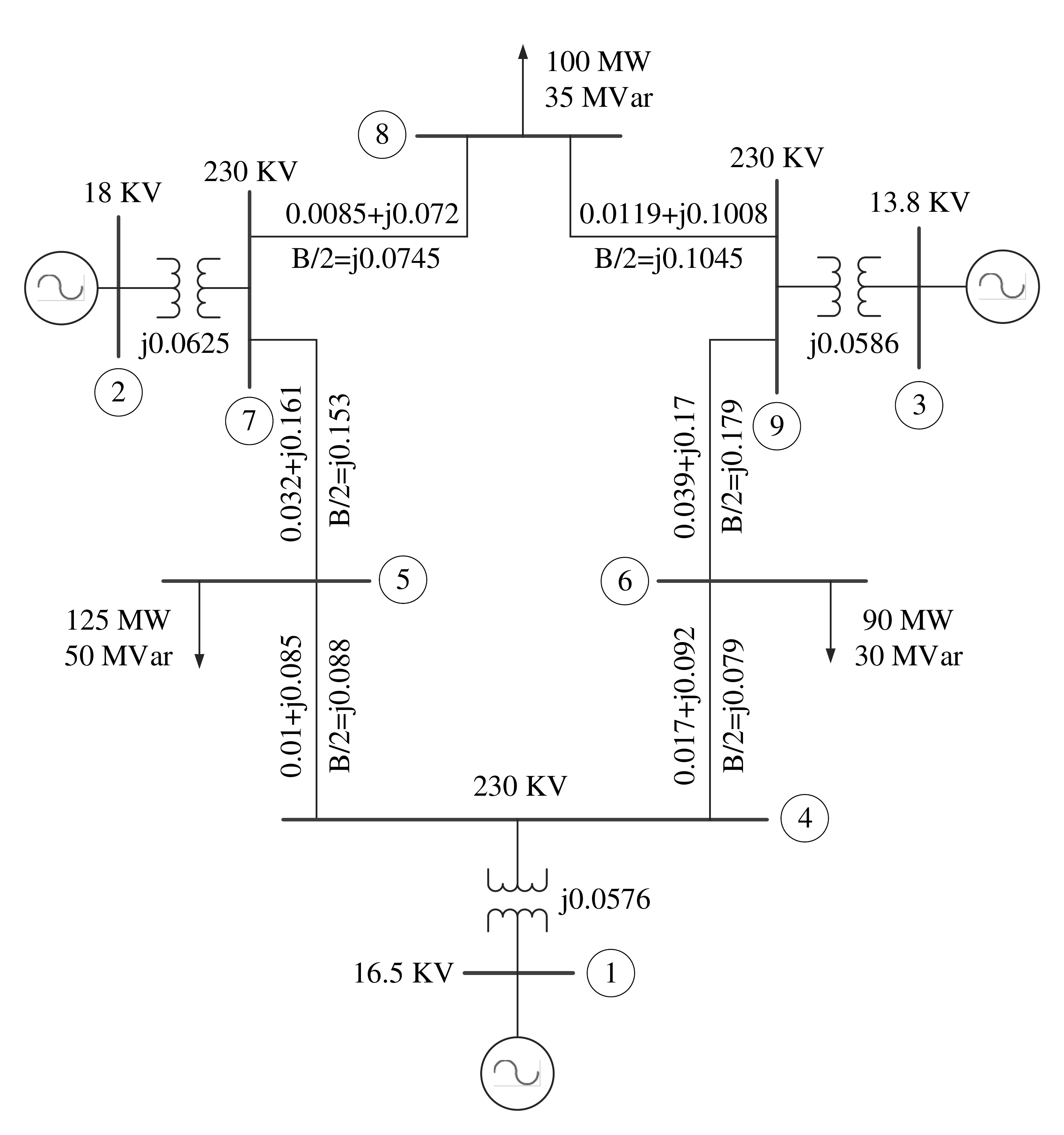}
    \caption{WECC-9 bus system}
    \label{fig:WECC}
\end{figure}

This system consists of three generators and three load points with total loading of 315 MW and 115 MVar as shown in \ref{fig:WECC}. System data and configuration are provided in \cite{AndersonFaud77} and inertia constants are given in Table \ref{tab:WECC}. This system has been extensively used in several power system stability studies.

\begin{table}[]
\caption{Machine inertia data for WECC 9-bus system\vspace{-1.5ex}}
\centering
\begin{tabular}{c|c|c|c}
\hline
Generator & H (p.u.)& D (p.u.) & $X_d$ (p.u.)  \T\\ \hline \hline
1                       & 23.64                         & $0.0255$& $0.0608$\T\\ \hline
2                       & 6.4                           & $0.00663$ & $0.1198$\T\\ \hline
3                       & 3.01                          & $0.00265$ & $0.1813$\T\\ \hline
\end{tabular}
\label{tab:WECC}
\end{table}
Simulation results for this case with fault near bus $8$ at time $1$ and the line $8$ -- $9$ is cleared after $2$ cycle. 
\subsubsection{Results with EKF}
\begin{figure}
    \hspace{-7ex}
    \includegraphics[scale=0.66]{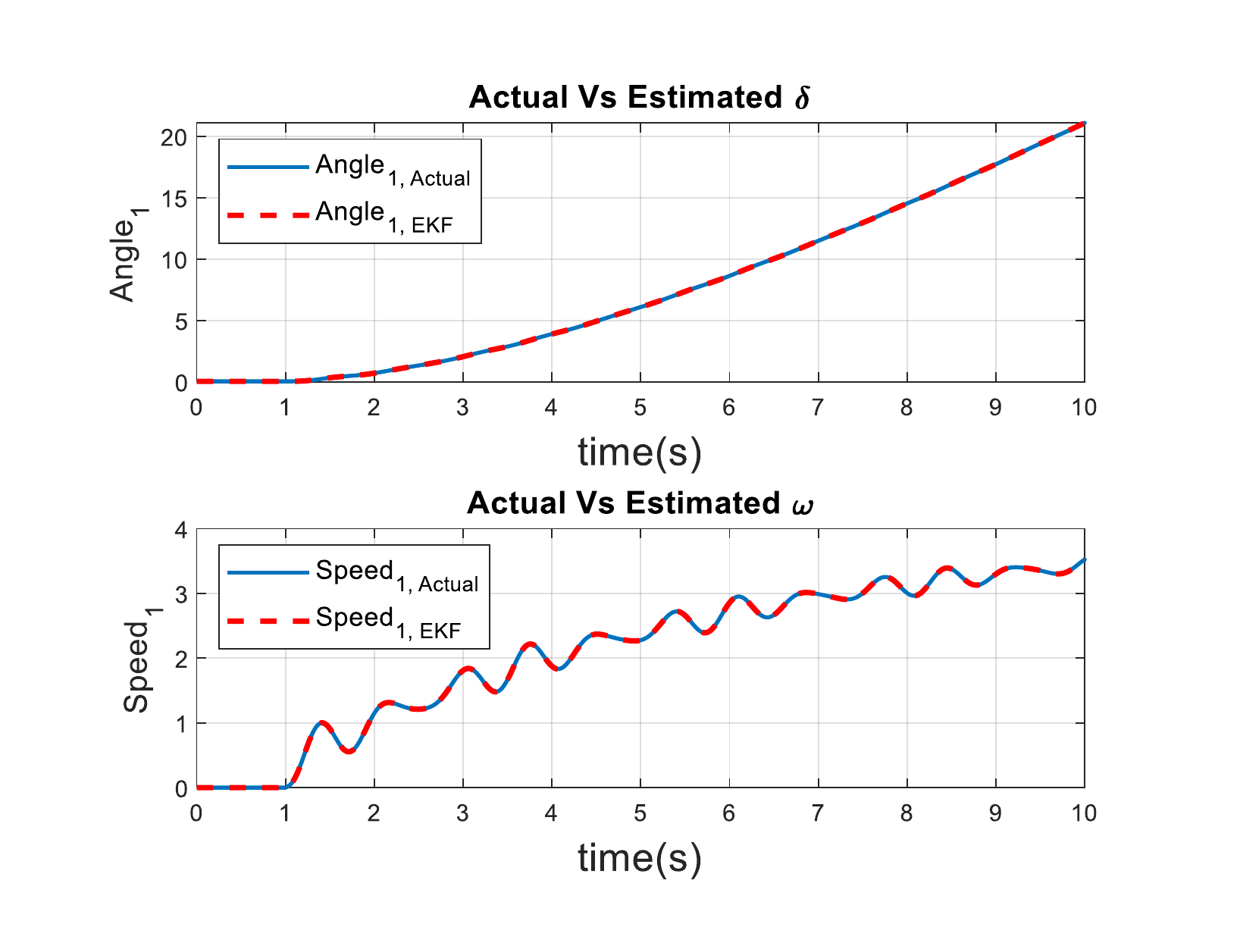}
    \caption{Actual vs Estimated Machine rotor angle and speed of generator 1 with EKF}
    \label{fig:WECC_Gen1_EKF}
\end{figure}

\begin{figure}
    \hspace{-7ex}
    \includegraphics[scale=0.66]{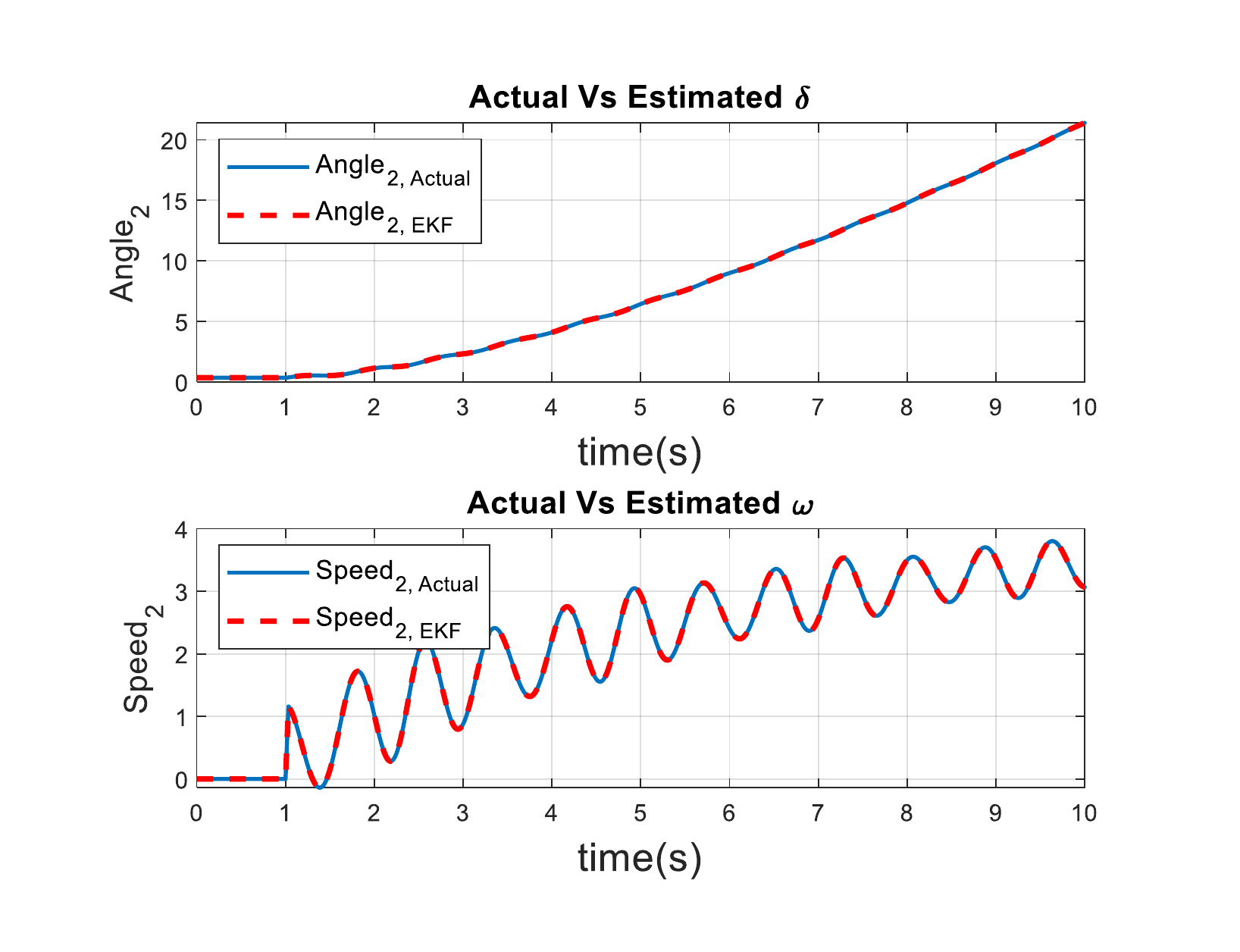}
    \caption{Actual vs Estimated Machine rotor angle and speed of generator 2 with EKF}
    \label{fig:WECC_Gen2_EKF}
\end{figure}

\begin{figure}
    \hspace{-7ex}
    \includegraphics[scale=0.66]{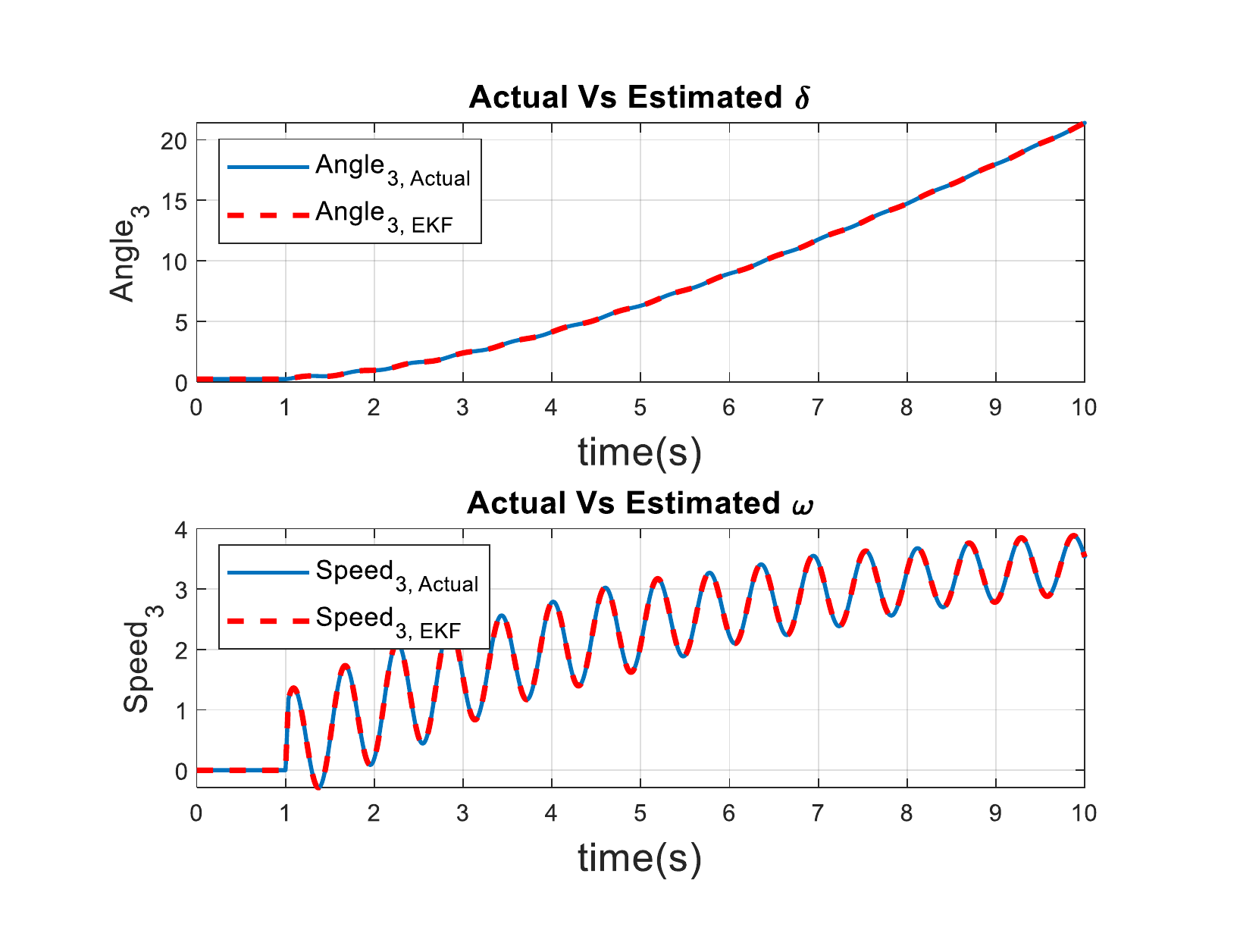}
    \caption{Actual vs Estimated Machine rotor angle and speed of generator 3 with EKF}
    \label{fig:WECC_Gen3_EKF}
\end{figure}

Figure~\ref{fig:WECC_Gen1_EKF}, Figure~\ref{fig:WECC_Gen2_EKF}, and Figure~\ref{fig:WECC_Gen3_EKF} show the plot of actual and estimated states (rotor angle and speed) of each generator in WECC 3-machine 9-bus system using EKF.

\subsubsection{Results With UKF}
\begin{figure}
    \hspace{-7ex}
    \includegraphics[scale=0.66]{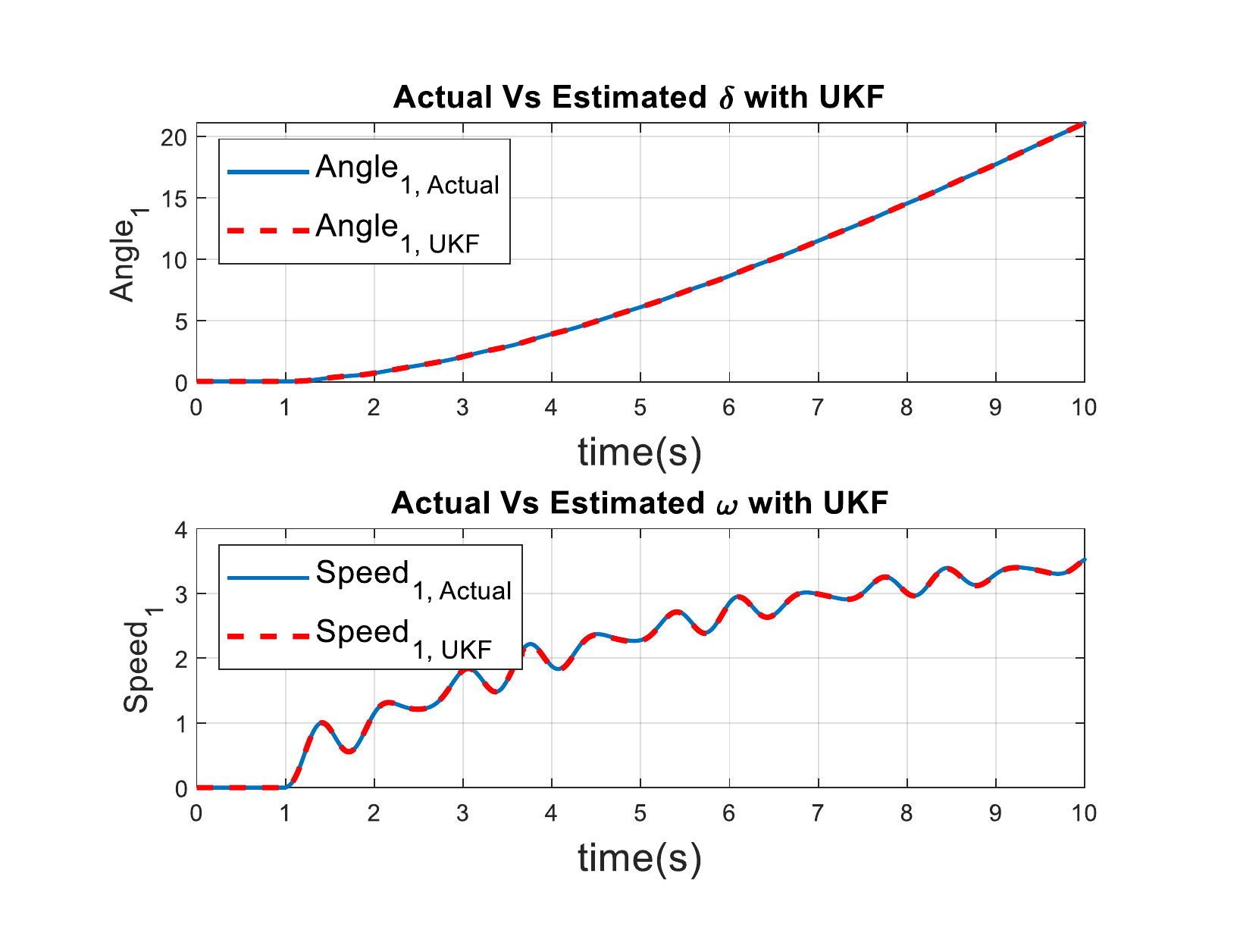}
    \caption{Actual vs Estimated Machine rotor angle and speed of generator 1 with UKF}
    \label{fig:WECC_Gen1_UKF}
\end{figure}

\begin{figure}
    \hspace{-7ex}
    \includegraphics[scale=0.66]{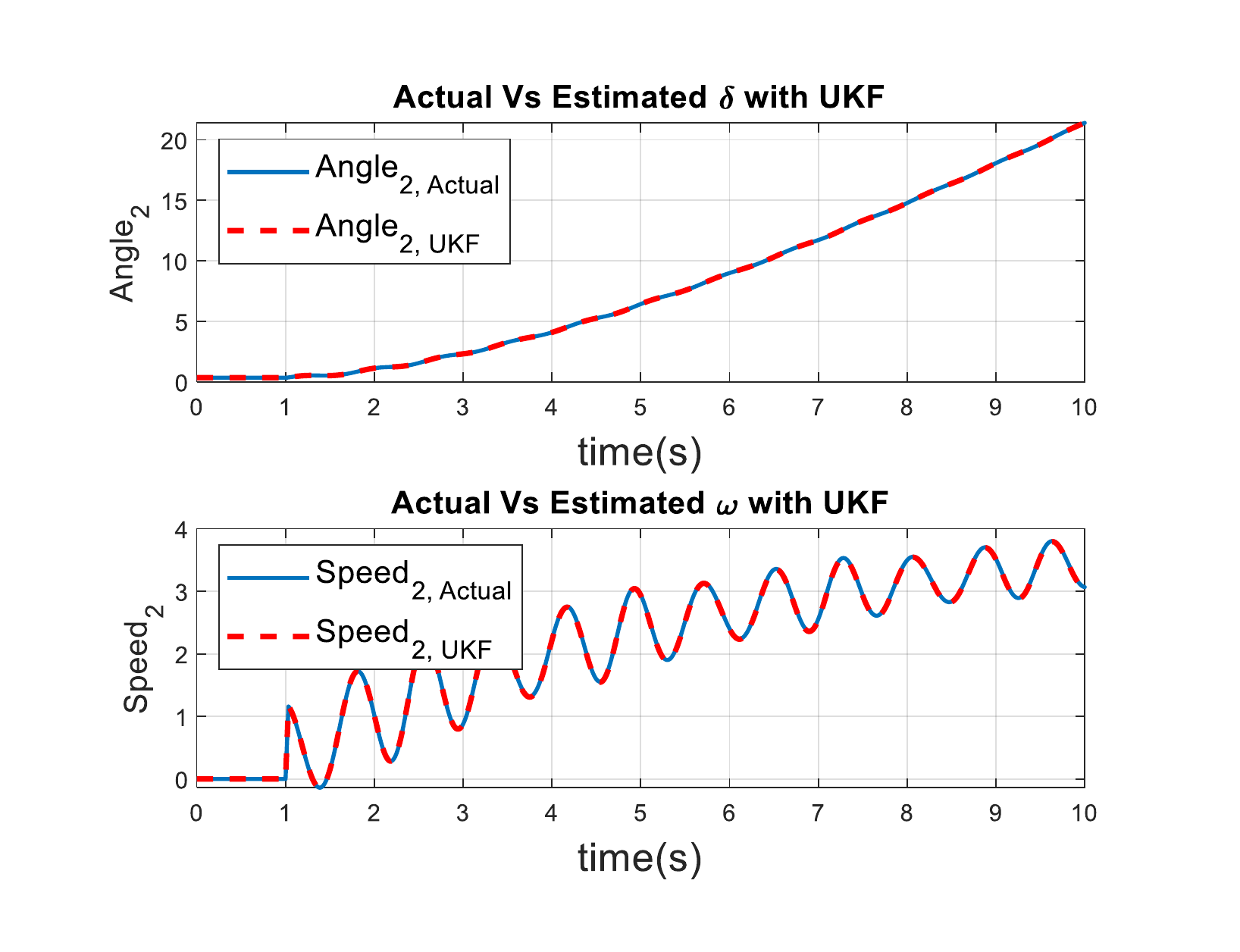}
    \caption{Actual vs Estimated Machine rotor angle and speed of generator 2 with UKF}
    \label{fig:WECC_Gen2_UKF}
\end{figure}

\begin{figure}
    \hspace{-7ex}
    \includegraphics[scale=0.66]{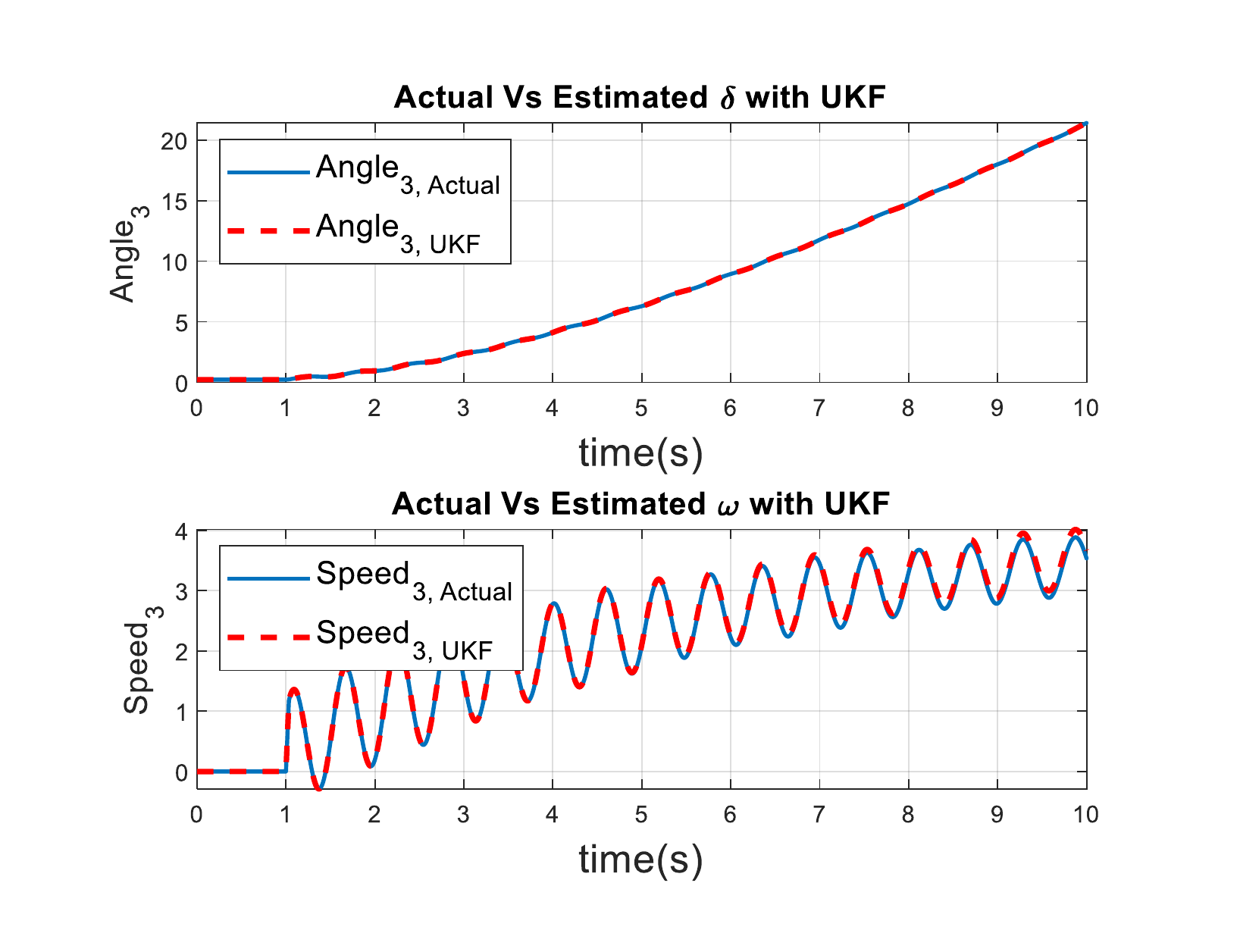}
    \caption{Actual vs Estimated Machine rotor angle and speed of generator 3 with UKF}
    \label{fig:WECC_Gen3_UKF}
\end{figure}

Figure~\ref{fig:WECC_Gen1_UKF}, Figure~\ref{fig:WECC_Gen2_UKF}, and Figure~\ref{fig:WECC_Gen3_UKF} show the plot of actual and estimated states (rotor angle and speed) of each generator in WECC 3-machine 9-bus system using UKF.

\subsection{Case New England 39-bus System}
 New England 39 bus system is characterized by ten generators, 21 load points with total loading of 6254.2 MW and 1387.1 MVar, the detail description of this system is provided in \cite{192898}. These systems have been tested for several studies on power system stability. The machine data used in the simulations are given in Table~\ref{tab:New_England39}. For this case study we have assumed that a fault occurs near bus $4$ after $1$ second and the fault is cleared after $2$ cycle by removing the line connecting bus $4$ and bus $14$. 
 

\begin{table}[]
\caption{Machine inertia data for New England-39 bus system\vspace{-1.5ex}}
\centering
\begin{tabular}{c|c|c|c}
\hline
Generator & H (p.u.)& $X_d$ (p.u.) & D (p.u.)  \T\\ \hline \hline
1                       & 500                         & $0.006$& $0$\T\\ \hline
2                       & 30.3                           & $0.0697$ & $0$\T\\ \hline
3                       & 35.8                          & $0.0531$ & $0$\T\\ \hline

4                       & 29.6                          & $0.0436$ & $0$\T\\ \hline

5                       & 26                          & $0.132$ & $0$\T\\ \hline

6                       & 34.8                          & $0.05$ & $0$\T\\ \hline
7                       & 26.4                          & $0.049$ & $0$\T\\ \hline
8                       & 24.3                          & $0.057$ & $0$\T\\ \hline
9                       & 34.5                          & $0.057$ & $0$\T\\ \hline
10                       & 42                          & $0.031$ & $0$\T\\ \hline

\end{tabular}
\label{tab:New_England39}
\end{table}

\subsubsection{Results with EKF}
\begin{figure}
    \hspace{-7ex}
    \includegraphics[scale=0.66]{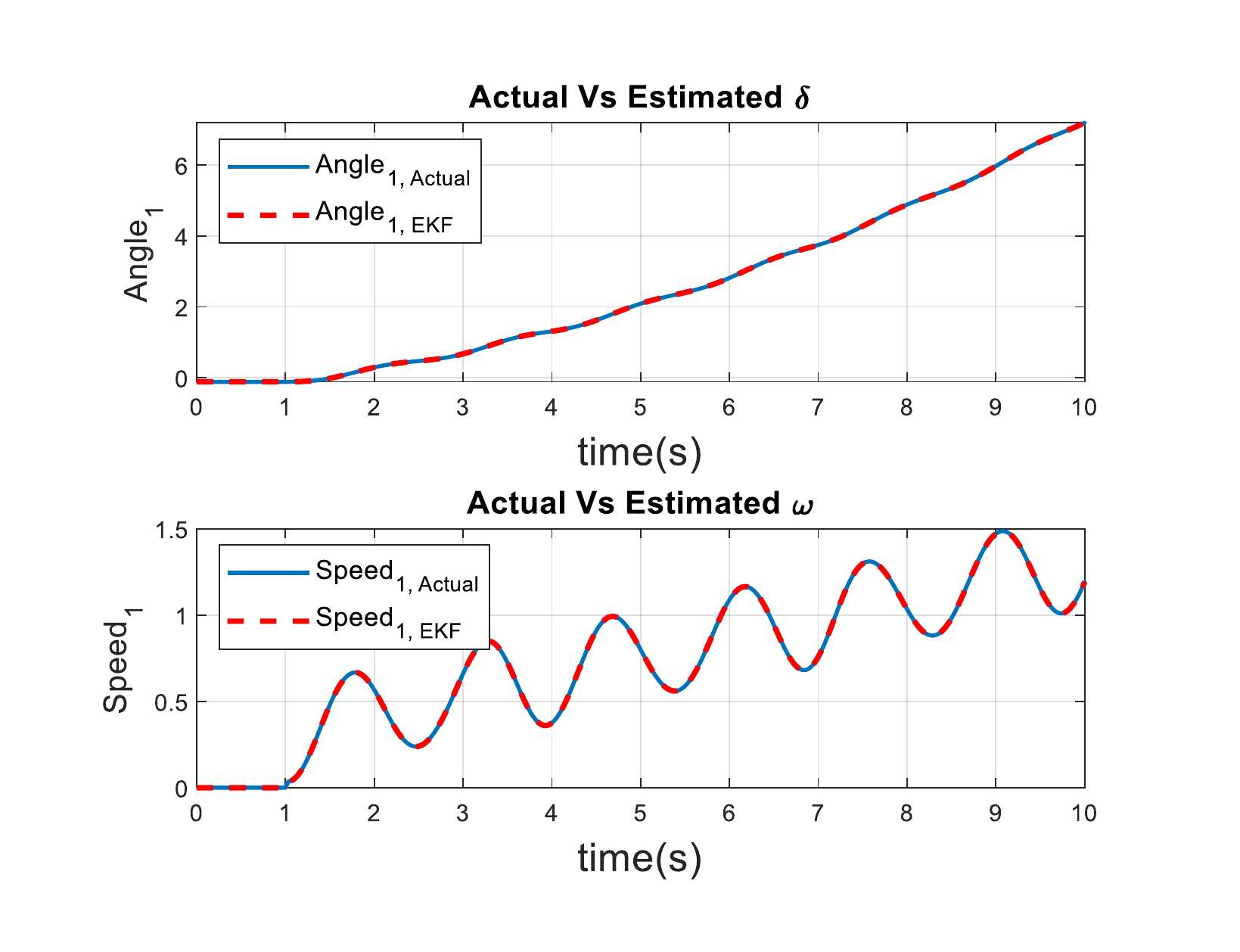}
    \caption{Actual vs estimated machine rotor angle and speed of generator 1 with EKF of New England 39 bus system.}
    \label{fig:39Generator_1_EKF}
\end{figure}

\begin{figure}
    \hspace{-7ex}
    \includegraphics[scale=0.66]{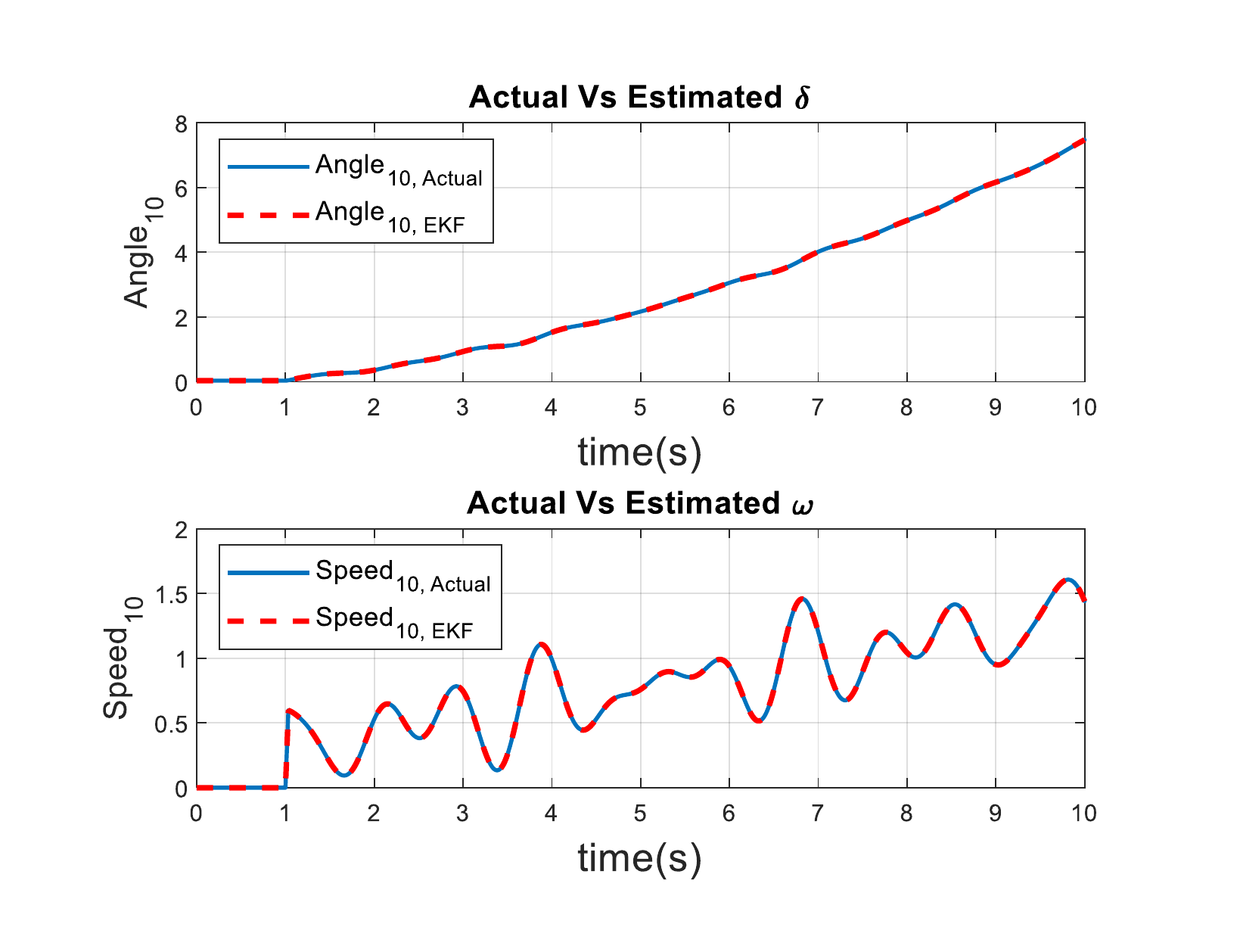}
    \caption{Actual vs estimated machine rotor angle and speed of generator 10 with EKF of New England 39 bus system.}
    \label{fig:39Generator_10_EKF}
\end{figure}

Figure~\ref{fig:39Generator_1_EKF}, to  Figure~\ref{fig:39Generator_10_EKF} show the plot of actual and estimated states (rotor angle and speed) of each generator in New England $10$-machine 39-bus system using EKF.

\begin{figure}
    \hspace{-7ex}
    \includegraphics[scale=0.66]{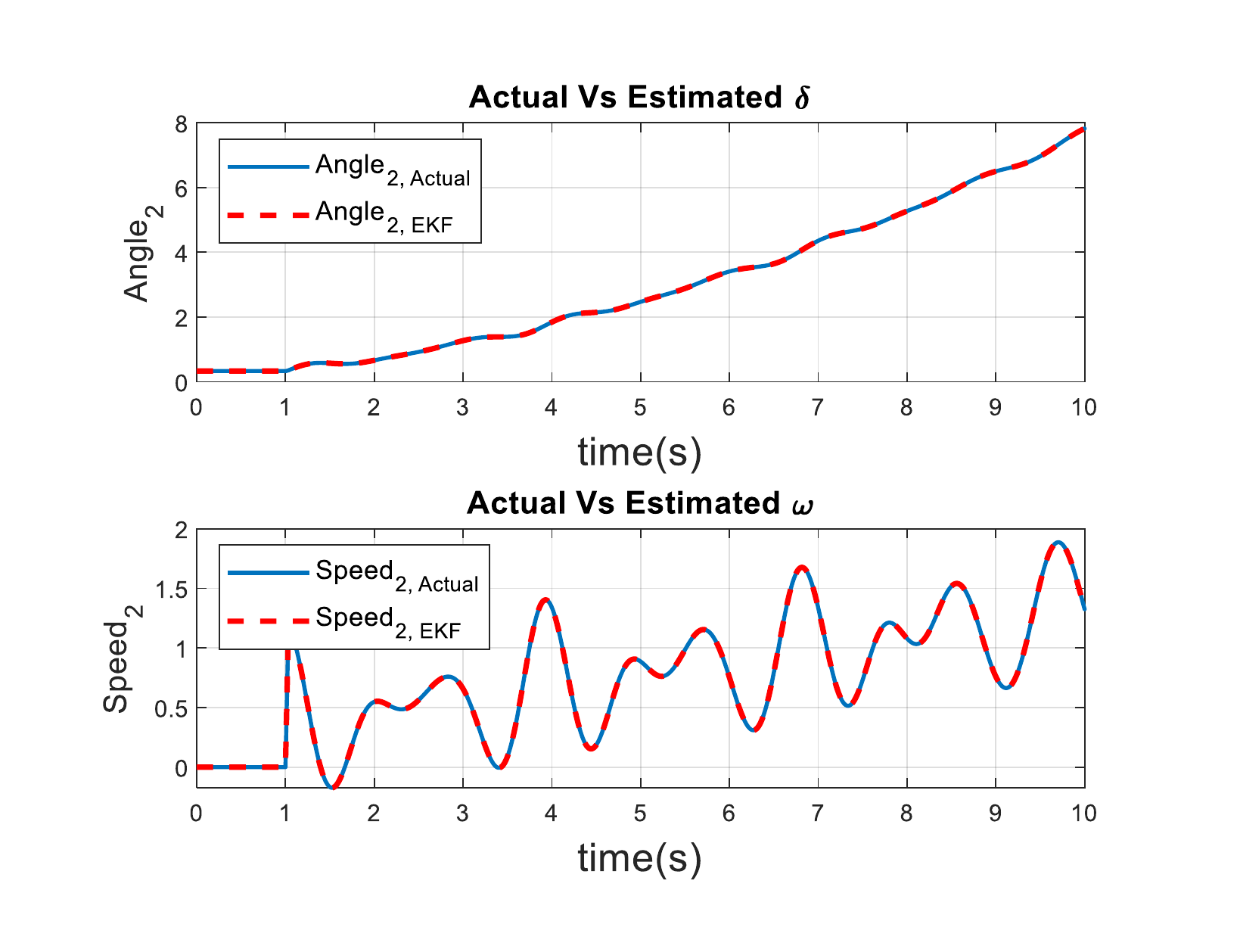}
    \caption{Actual vs estimated machine rotor angle and speed of generator 2 with EKF of New England 39 bus system.}
    \label{fig:39Generator_2_EKF}
\end{figure}

\begin{figure}
    \hspace{-7ex}
    \includegraphics[scale=0.66]{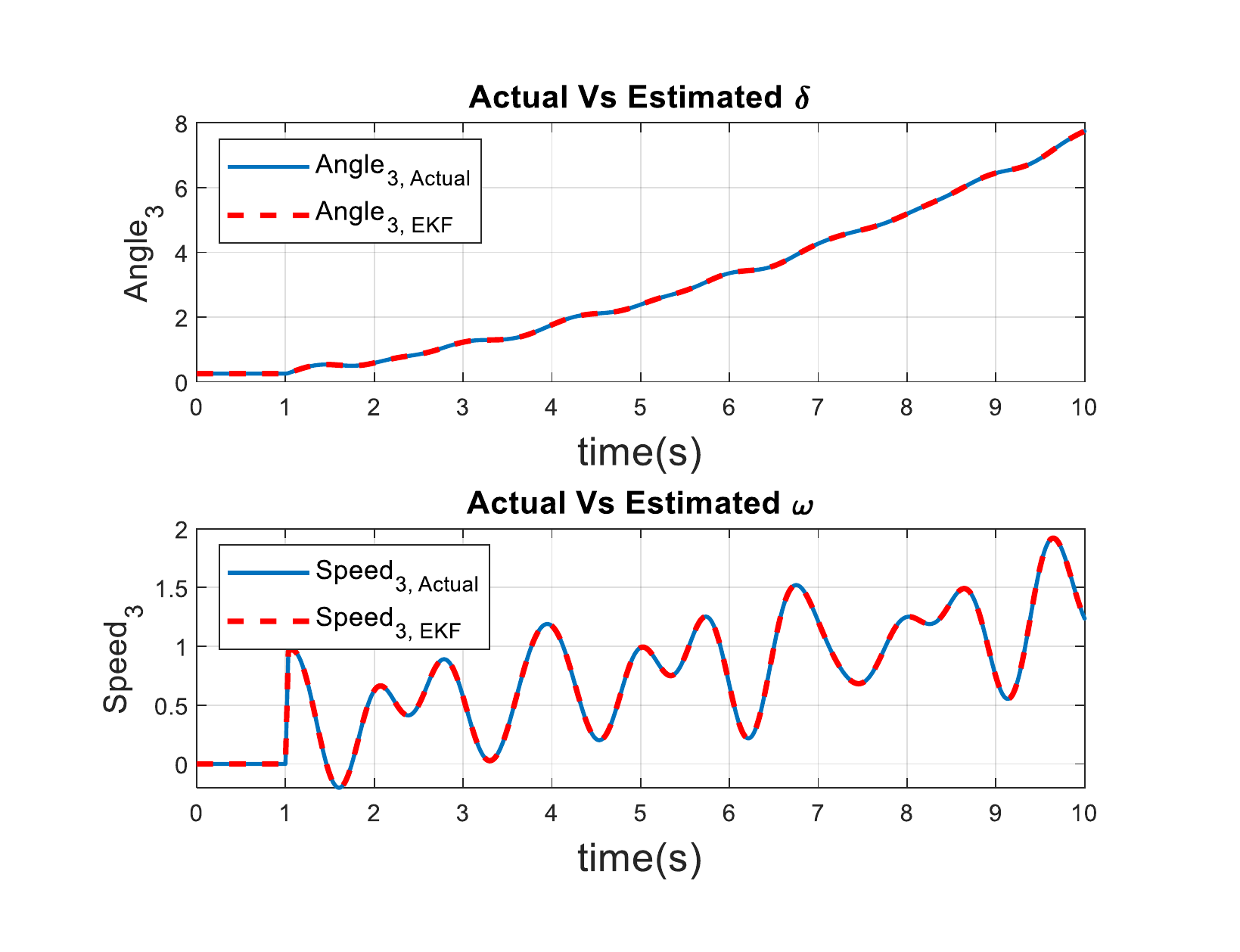}
    \caption{Actual vs estimated machine rotor angle and speed of generator 3 with EKF of New England 39 bus system.}
    \label{fig:39Generator_3_EKF}
\end{figure}

\begin{figure}
    \hspace{-7ex}
    \includegraphics[scale=0.66]{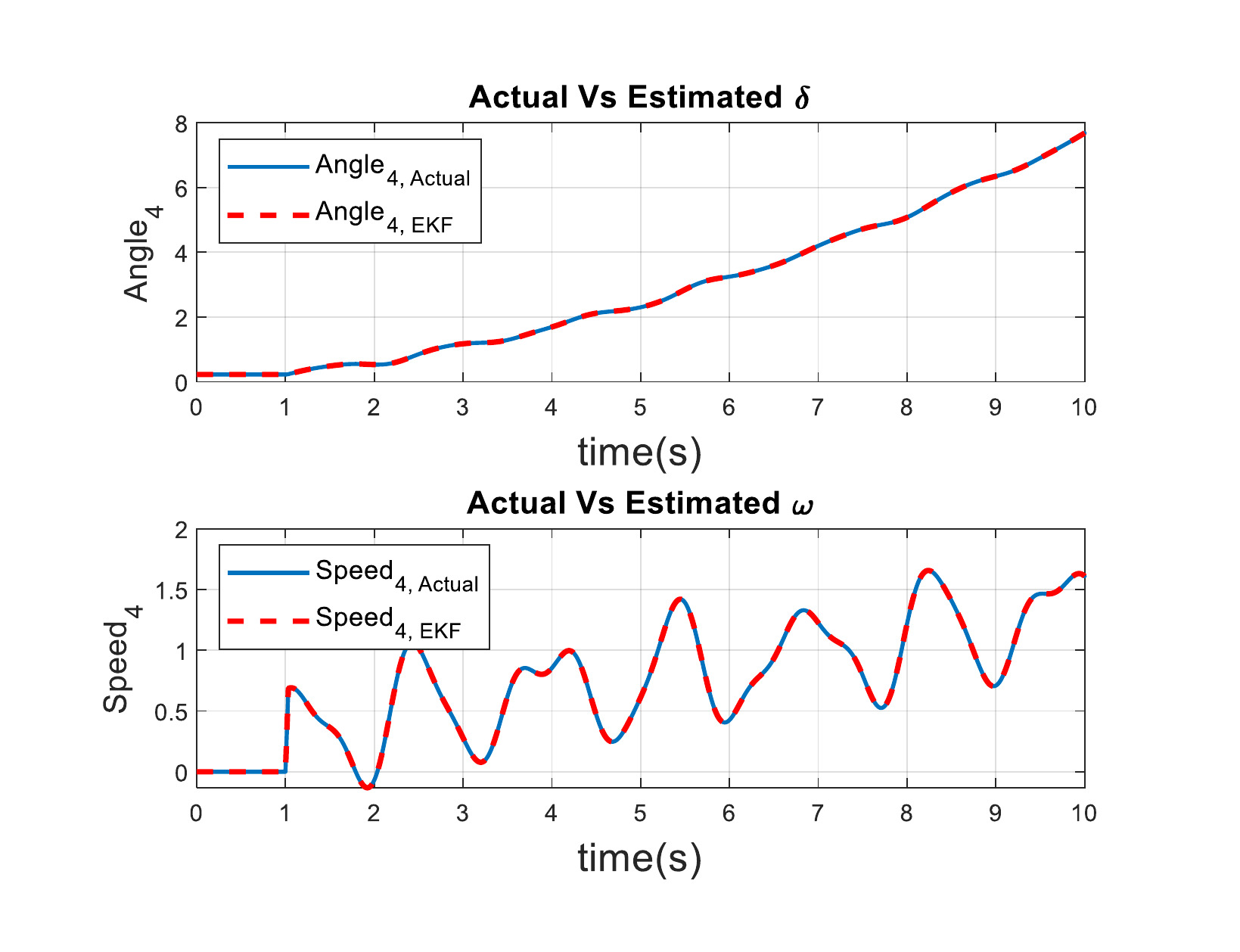}
    \caption{Actual vs estimated machine rotor angle and speed of generator 4 with EKF of New England 39 bus system.}
    \label{fig:39Generator_4_EKF}
\end{figure}

\begin{figure}
    \hspace{-7ex}
    \includegraphics[scale=0.66]{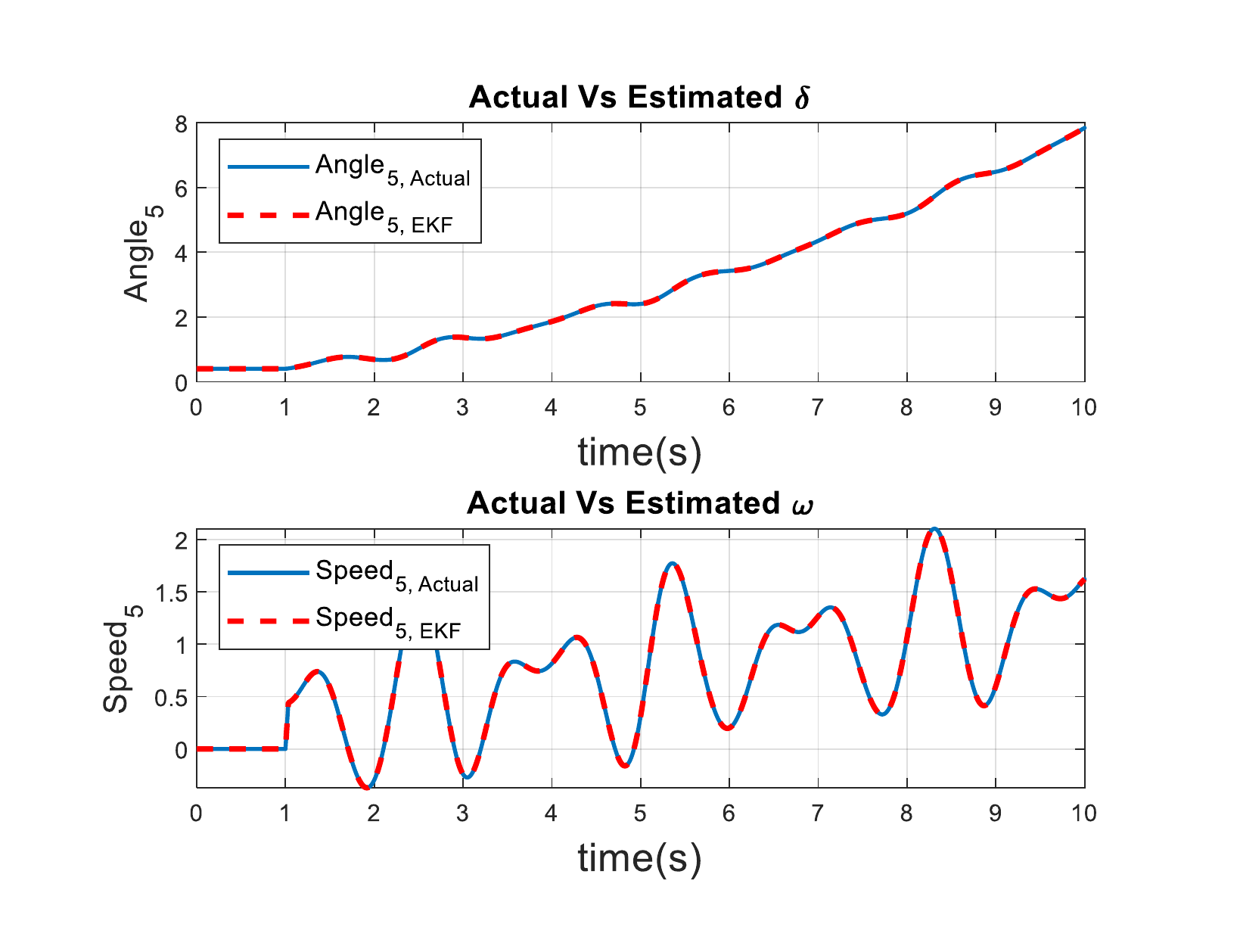}
    \caption{Actual vs estimated machine rotor angle and speed of generator 5 with EKF of New England 39 bus system.}
    \label{fig:39Generator_5_EKF}
\end{figure}

\begin{figure}
    \hspace{-7ex}
    \includegraphics[scale=0.66]{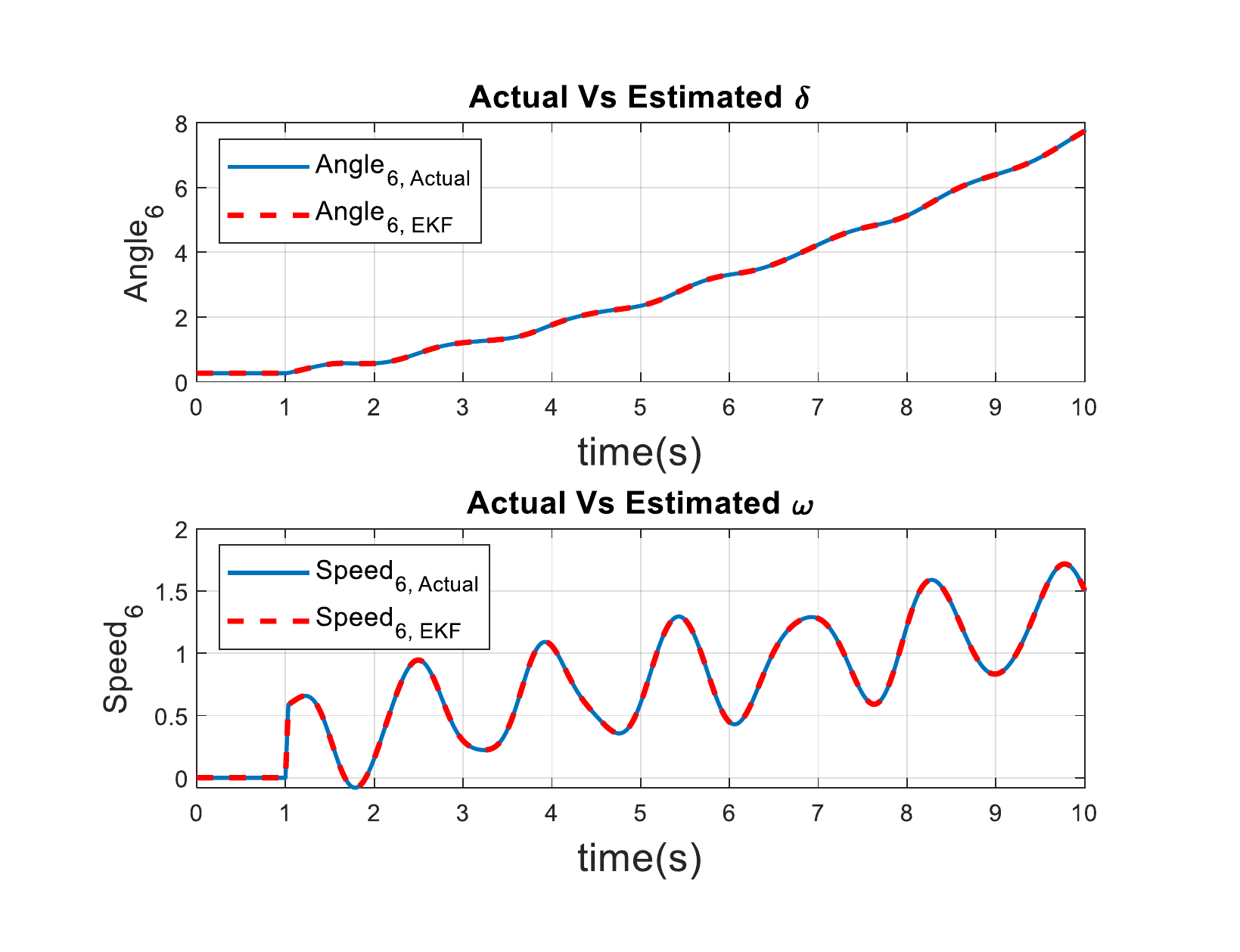}
    \caption{Actual vs estimated machine rotor angle and speed of generator 6 with EKF of New England 39 bus system.}
    \label{fig:39Generator_6_EKF}
\end{figure}

\begin{figure}
    \hspace{-7ex}
    \includegraphics[scale=0.66]{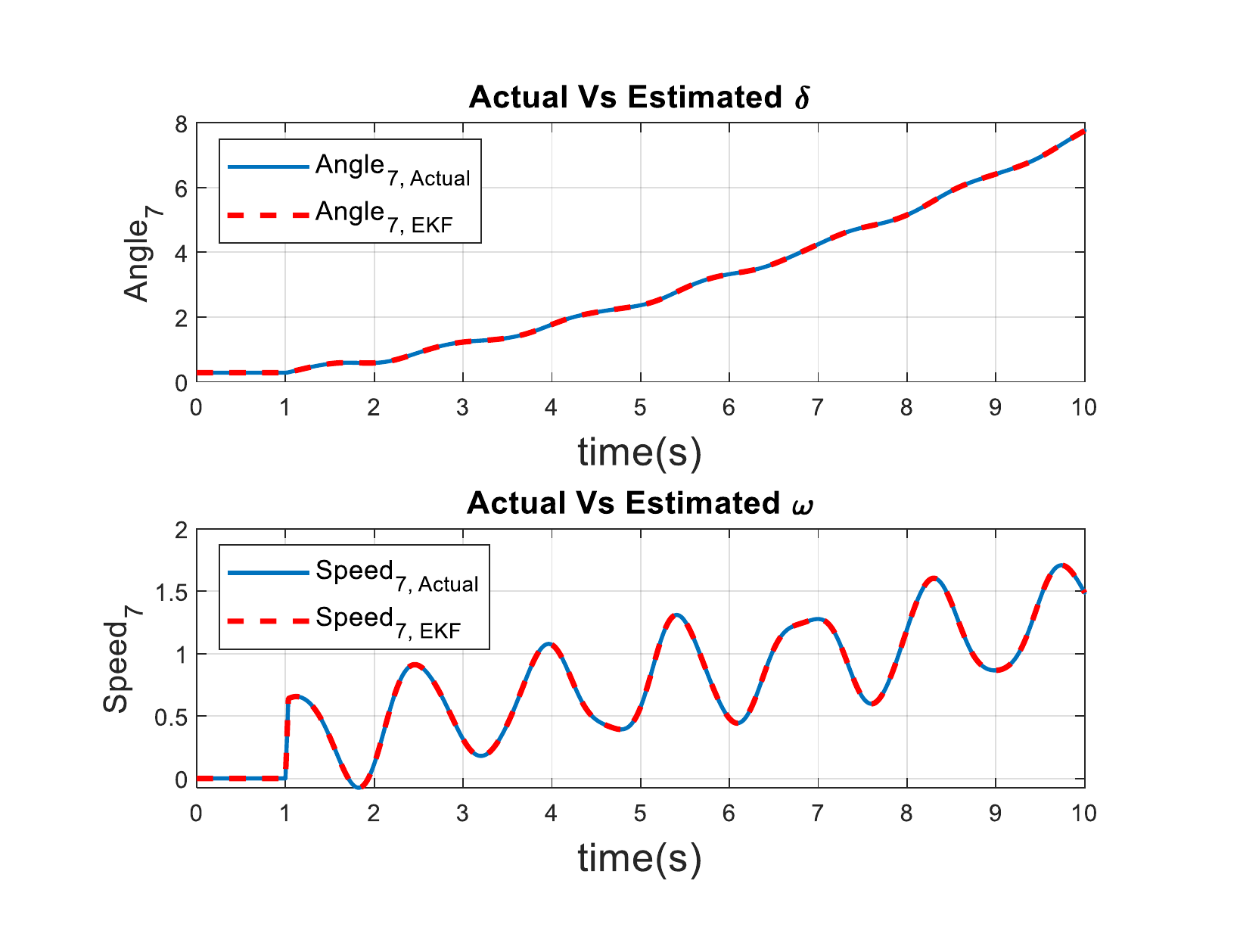}
    \caption{Actual vs estimated machine rotor angle and speed of generator 7 with EKF of New England 39 bus system.}
    \label{fig:39Generator_7_EKF}
\end{figure}

\begin{figure}
    \hspace{-7ex}
    \includegraphics[scale=0.66]{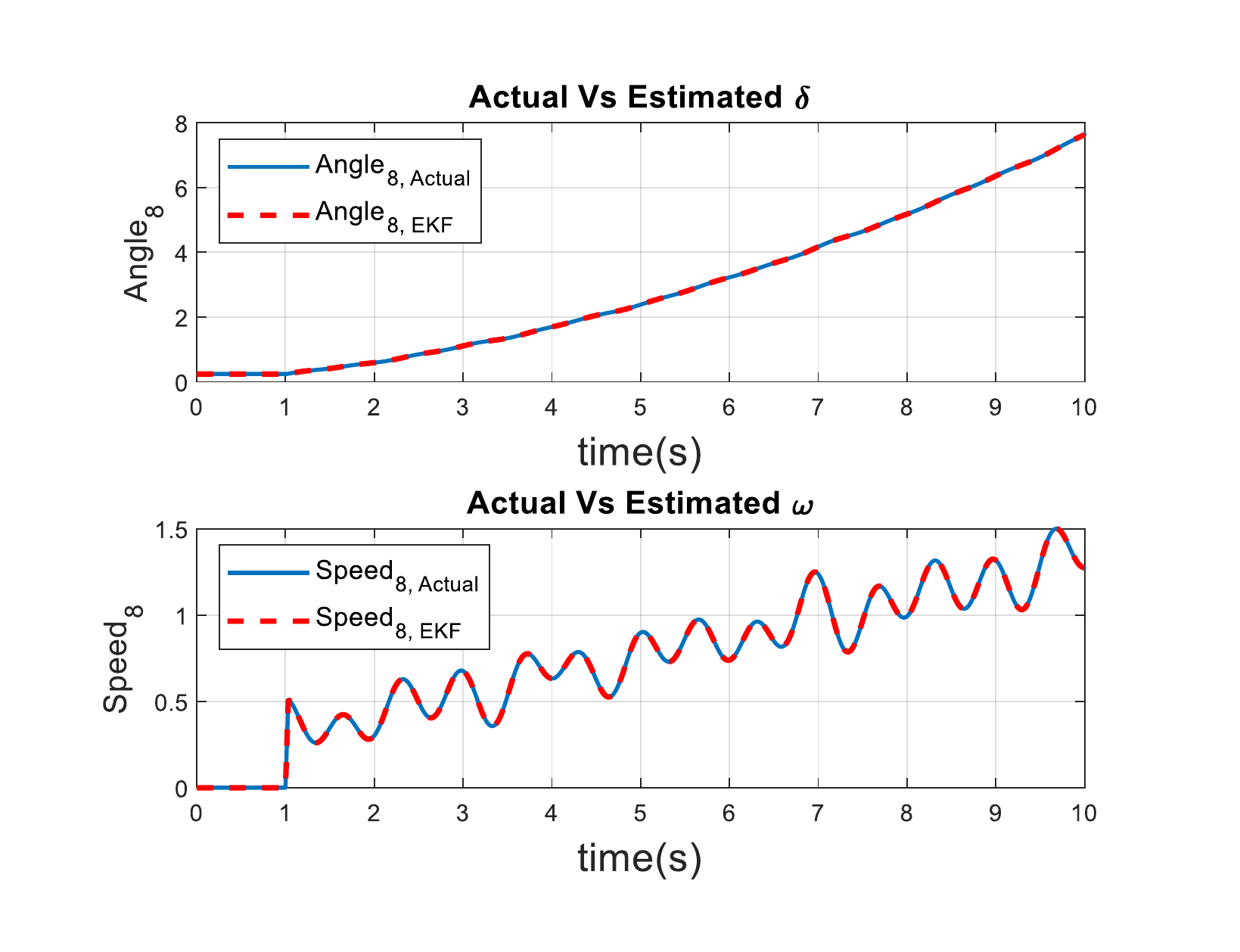}
    \caption{Actual vs estimated machine rotor angle and speed of generator 8 with EKF of New England 39 bus system.}
    \label{fig:39Generator_8_EKF}
\end{figure}

\begin{figure}
    \hspace{-7ex}
    \includegraphics[scale=0.66]{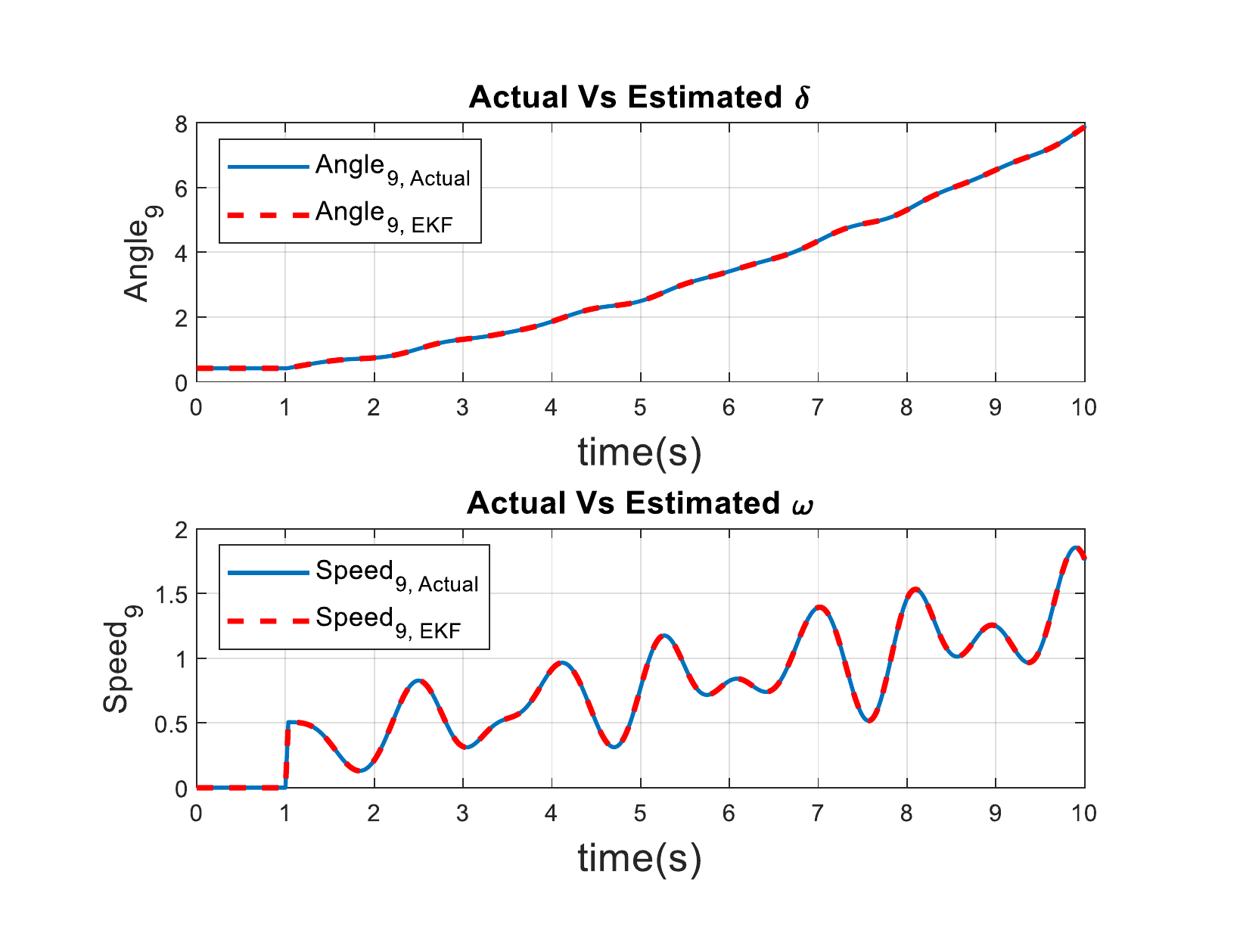}
    \caption{Actual vs estimated machine rotor angle and speed of generator 9 with EKF of New England 39 bus system.}
    \label{fig:39Generator_9_EKF}
\end{figure}

\subsubsection{Results with UKF}
\begin{figure}
    \hspace{-7ex}
    \includegraphics[scale=0.66]{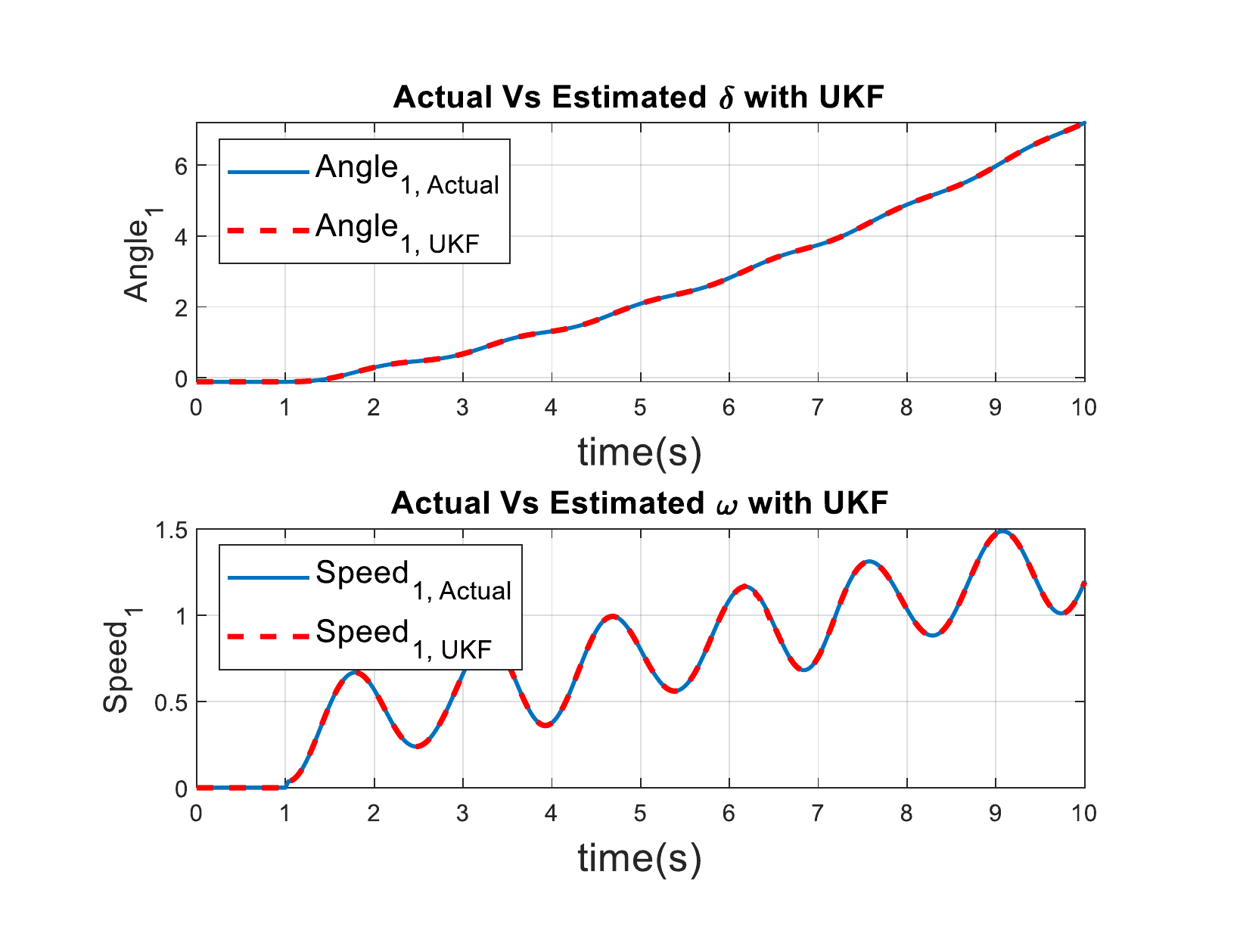}
    \caption{Actual vs estimated machine rotor angle and speed of generator 1 with UKF of New England 39 bus system.}
    \label{fig:39Generator_1_UKF}
\end{figure}

\begin{figure}
    \hspace{-7ex}
    \includegraphics[scale=0.66]{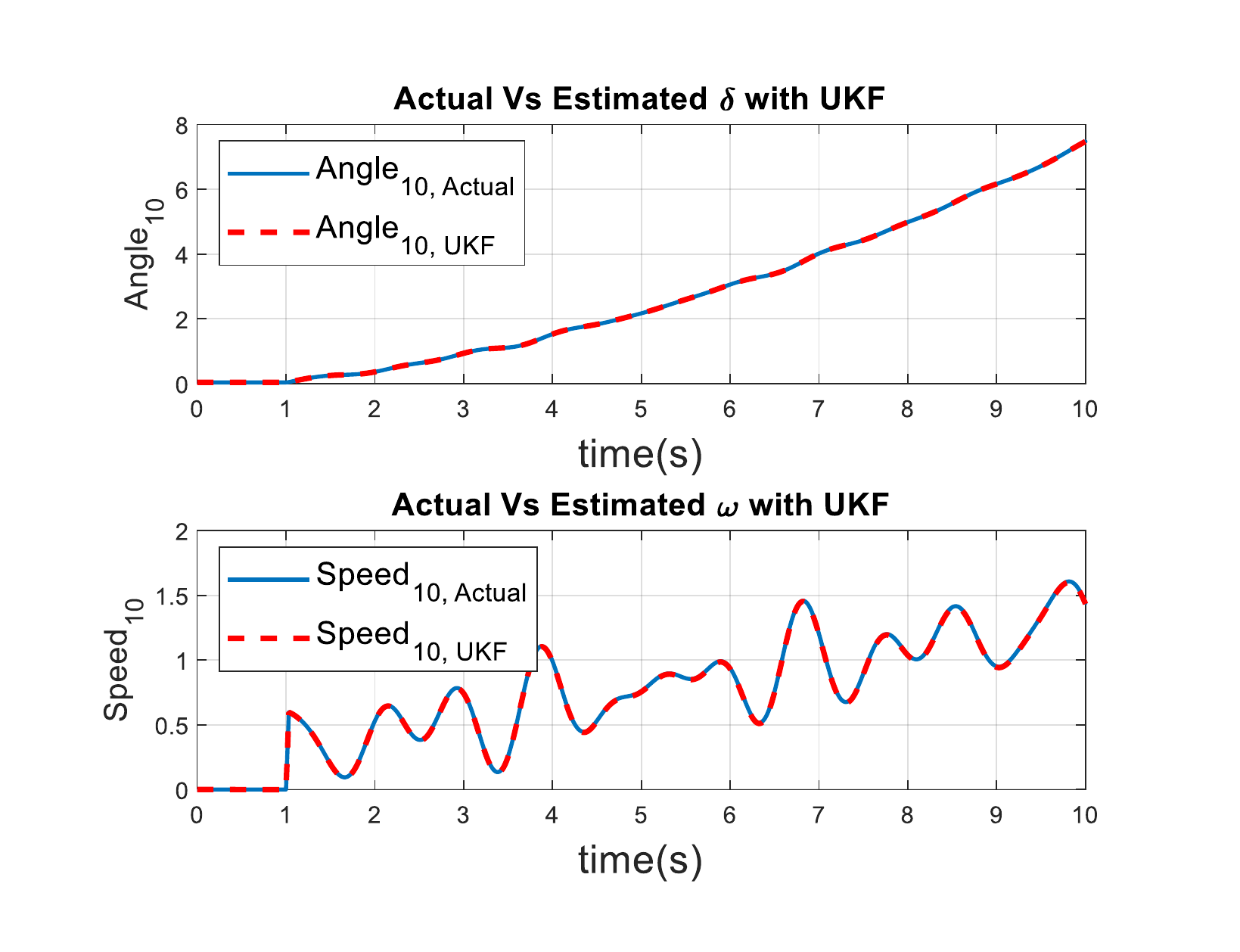}
    \caption{Actual vs estimated machine rotor angle and speed of generator 10 with UKF of New England 39 bus system.}
    \label{fig:39Generator_10_UKF}
\end{figure}

Figure~\ref{fig:39Generator_1_UKF}, to  Figure~\ref{fig:39Generator_10_UKF} show the plot of actual and estimated states (rotor angle and speed) of each generator in New England $10$-machine 39-bus system using UKF.

\begin{figure}
    \hspace{-7ex}
    \includegraphics[scale=0.66]{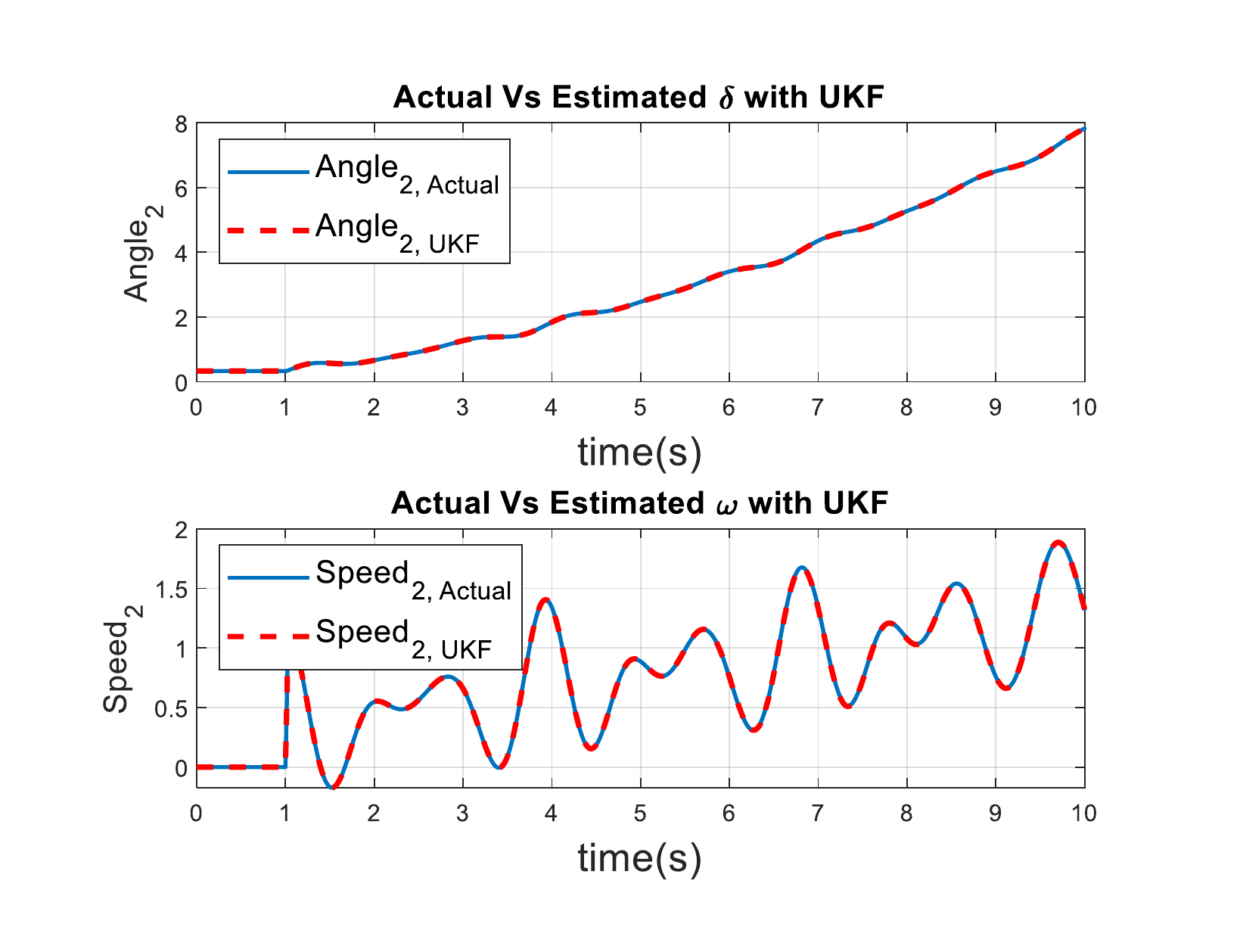}
    \caption{Actual vs estimated machine rotor angle and speed of generator 2 with UKF of New England 39 bus system.}
    \label{fig:39Generator_2_UKF}
\end{figure}

\begin{figure}
    \hspace{-7ex}
    \includegraphics[scale=0.66]{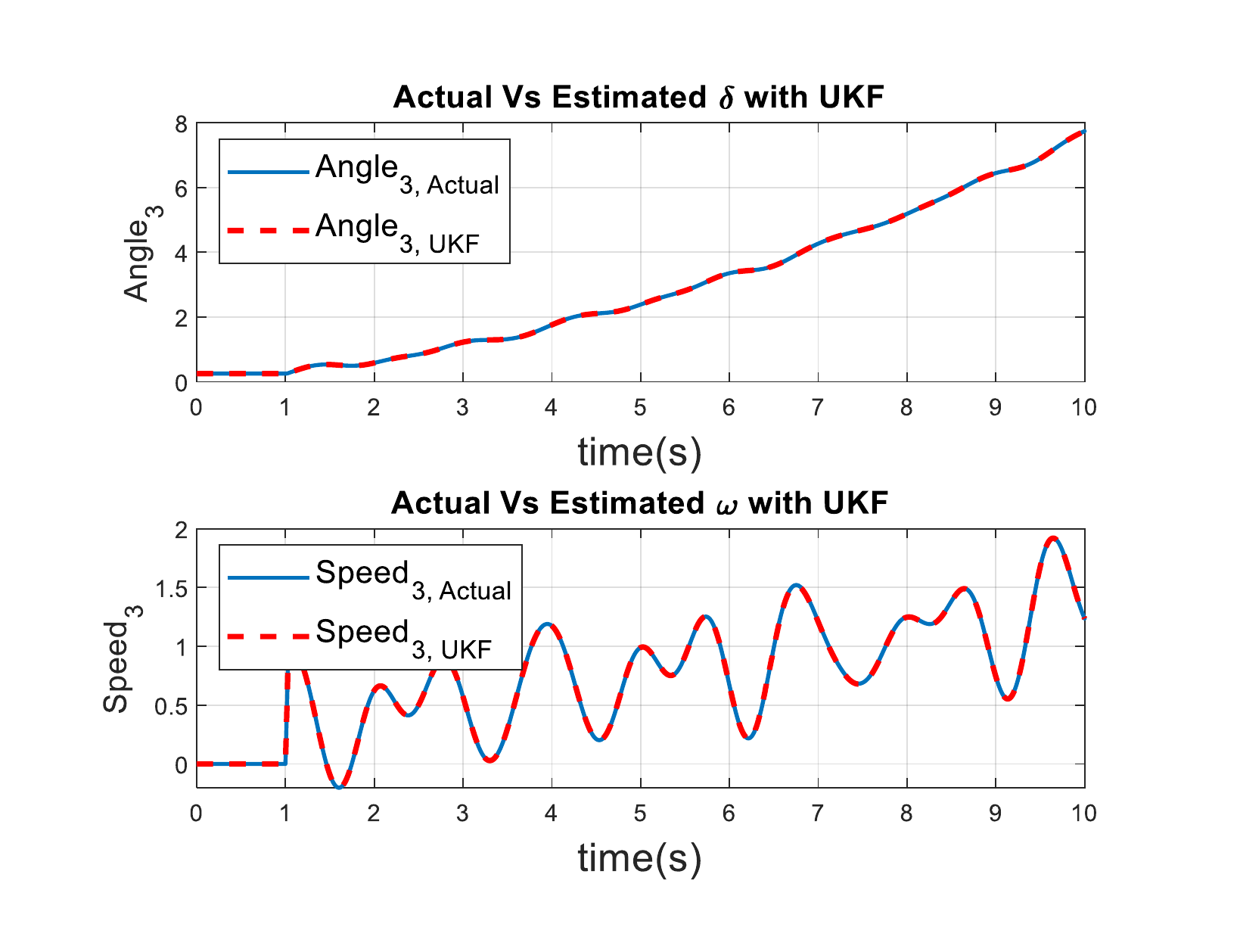}
    \caption{Actual vs estimated machine rotor angle and speed of generator 3 with UKF of New England 39 bus system.}
    \label{fig:39Generator_3_UKF}
\end{figure}

\begin{figure}
    \hspace{-7ex}
    \includegraphics[scale=0.66]{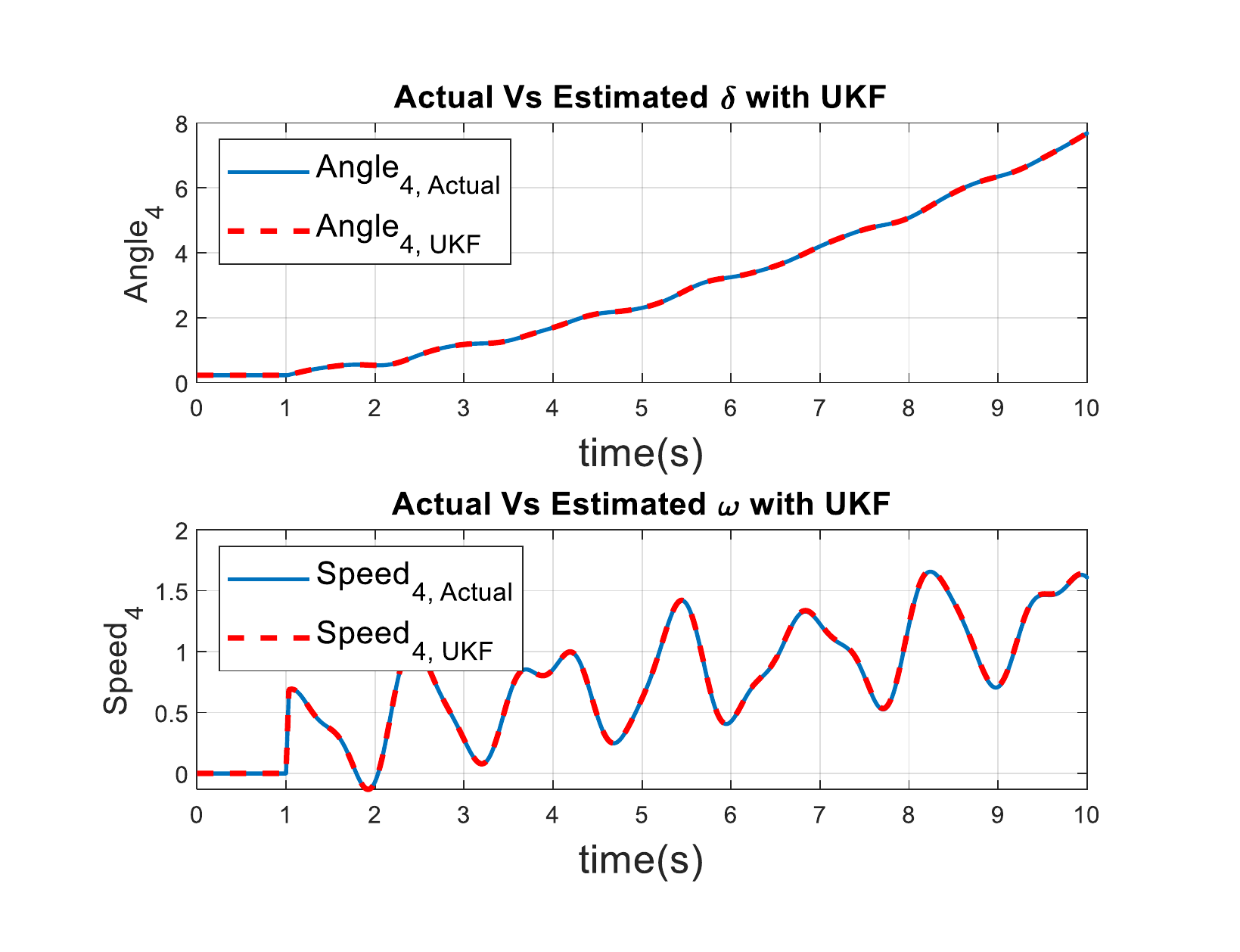}
    \caption{Actual vs estimated machine rotor angle and speed of generator 4 with UKF of New England 39 bus system.}
    \label{fig:39Generator_4_UKF}
\end{figure}

\begin{figure}
    \hspace{-7ex}
    \includegraphics[scale=0.66]{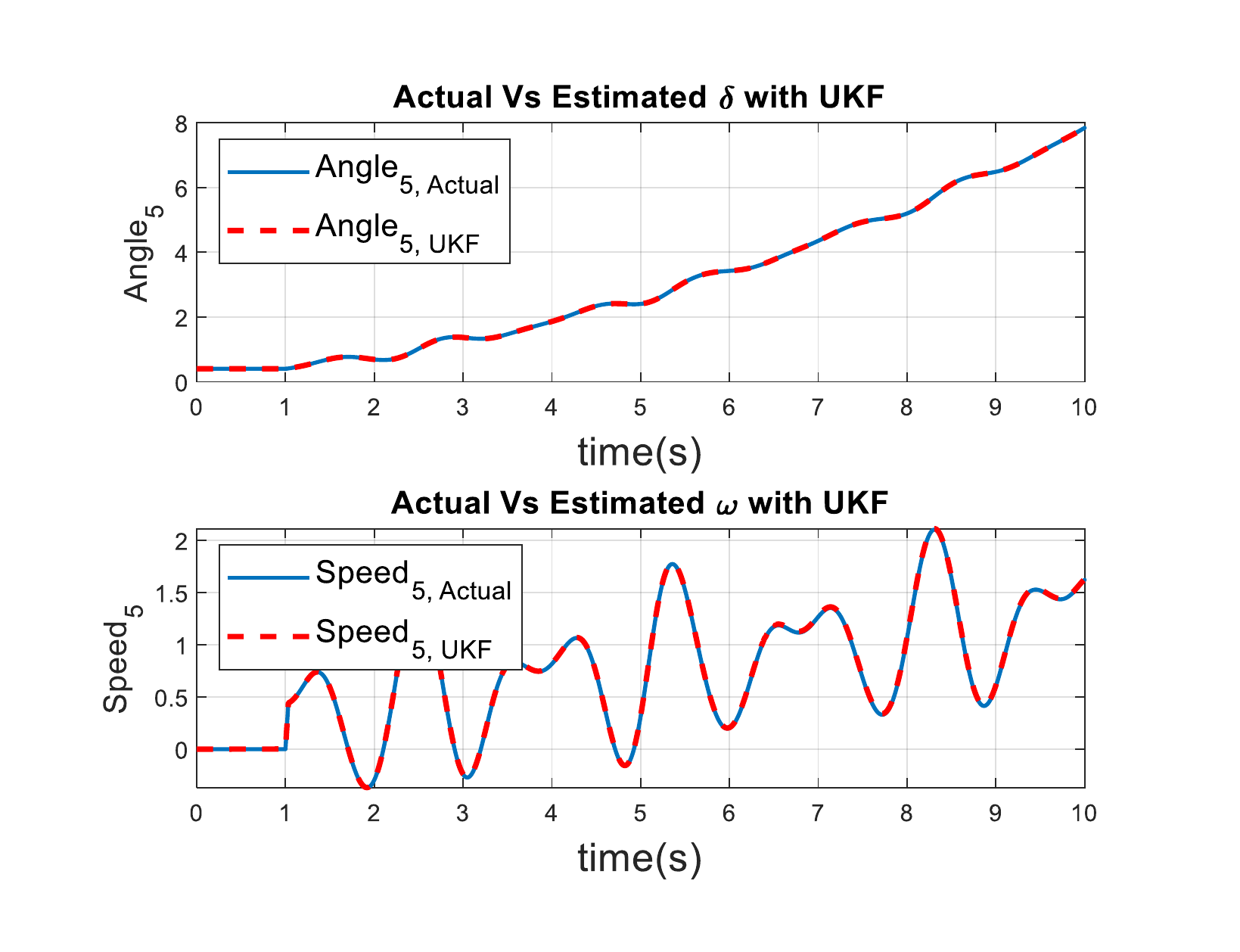}
    \caption{Actual vs estimated machine rotor angle and speed of generator 5 with UKF of New England 39 bus system.}
    \label{fig:39Generator_5_UKF}
\end{figure}

\begin{figure}
    \hspace{-7ex}
    \includegraphics[scale=0.66]{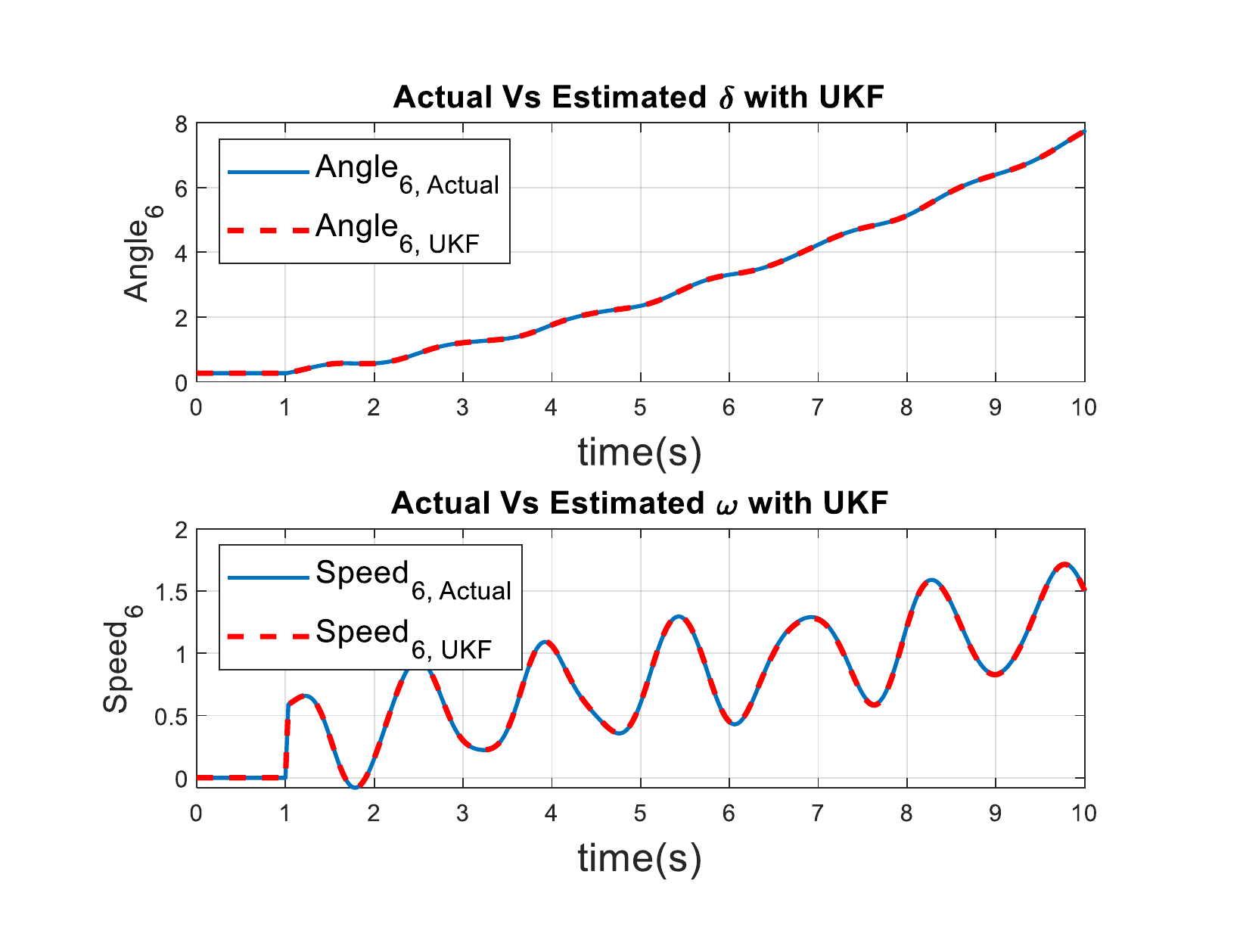}
    \caption{Actual vs estimated machine rotor angle and speed of generator 6 with UKF of New England 39 bus system.}
    \label{fig:39Generator_6_UKF}
\end{figure}

\begin{figure}
    \hspace{-7ex}
    \includegraphics[scale=0.66]{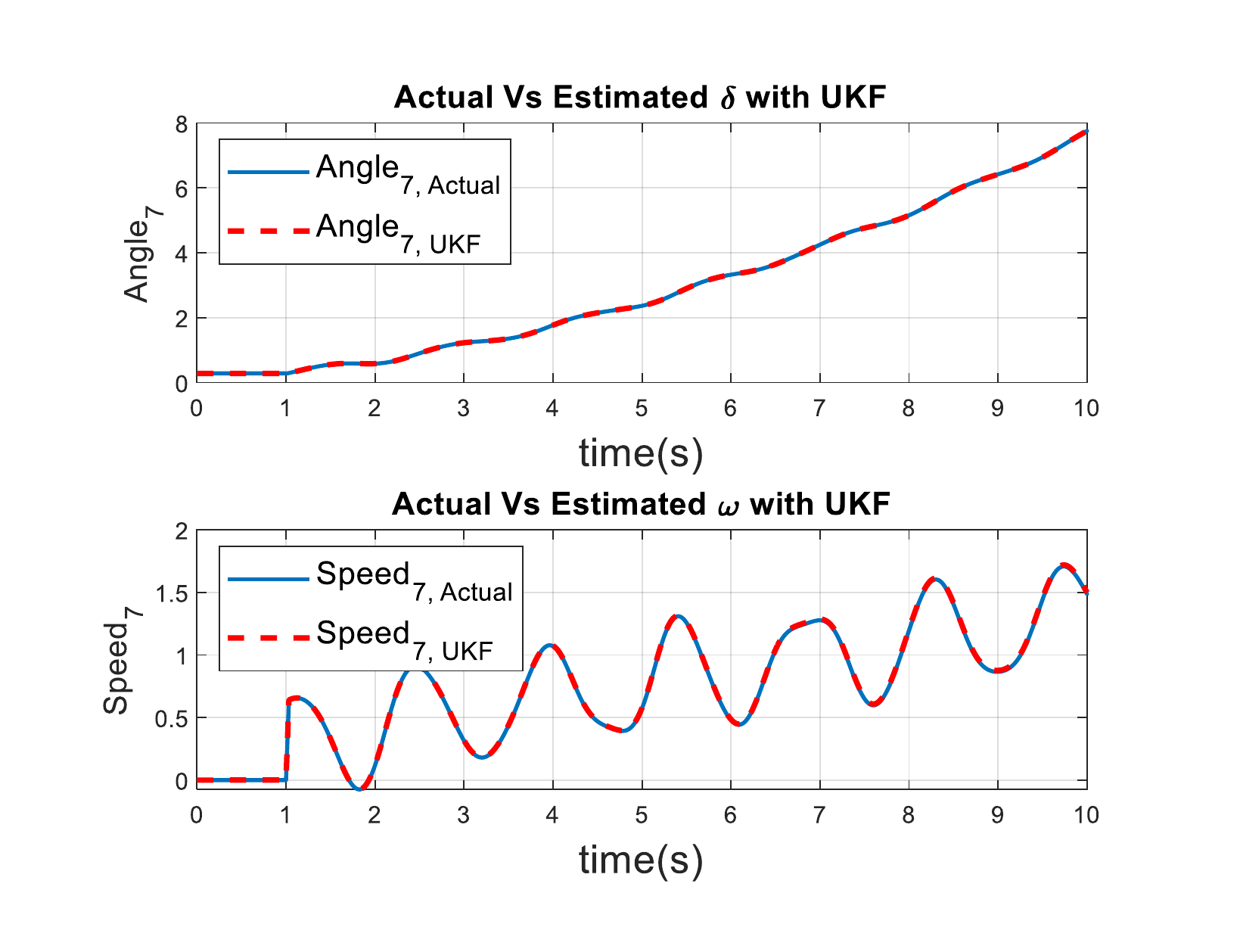}
    \caption{Actual vs estimated machine rotor angle and speed of generator 7 with UKF of New England 39 bus system.}
    \label{fig:39Generator_7_UKF}
\end{figure}

\begin{figure}
    \hspace{-7ex}
    \includegraphics[scale=0.66]{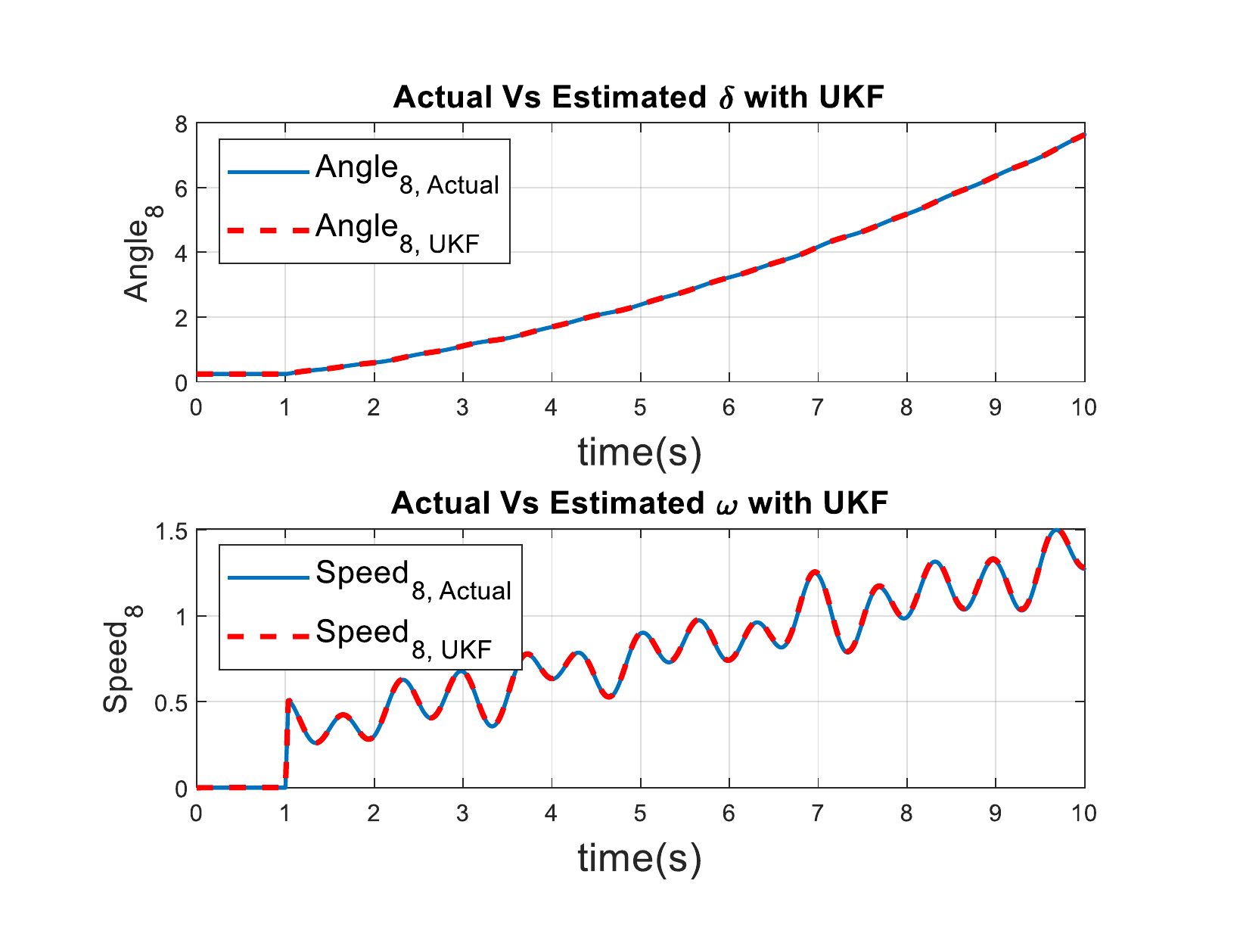}
    \caption{Actual vs estimated machine rotor angle and speed of generator 8 with UKF of New England 39 bus system.}
    \label{fig:39Generator_8_UKF}
\end{figure}

\begin{figure}
    \hspace{-7ex}
    \includegraphics[scale=0.66]{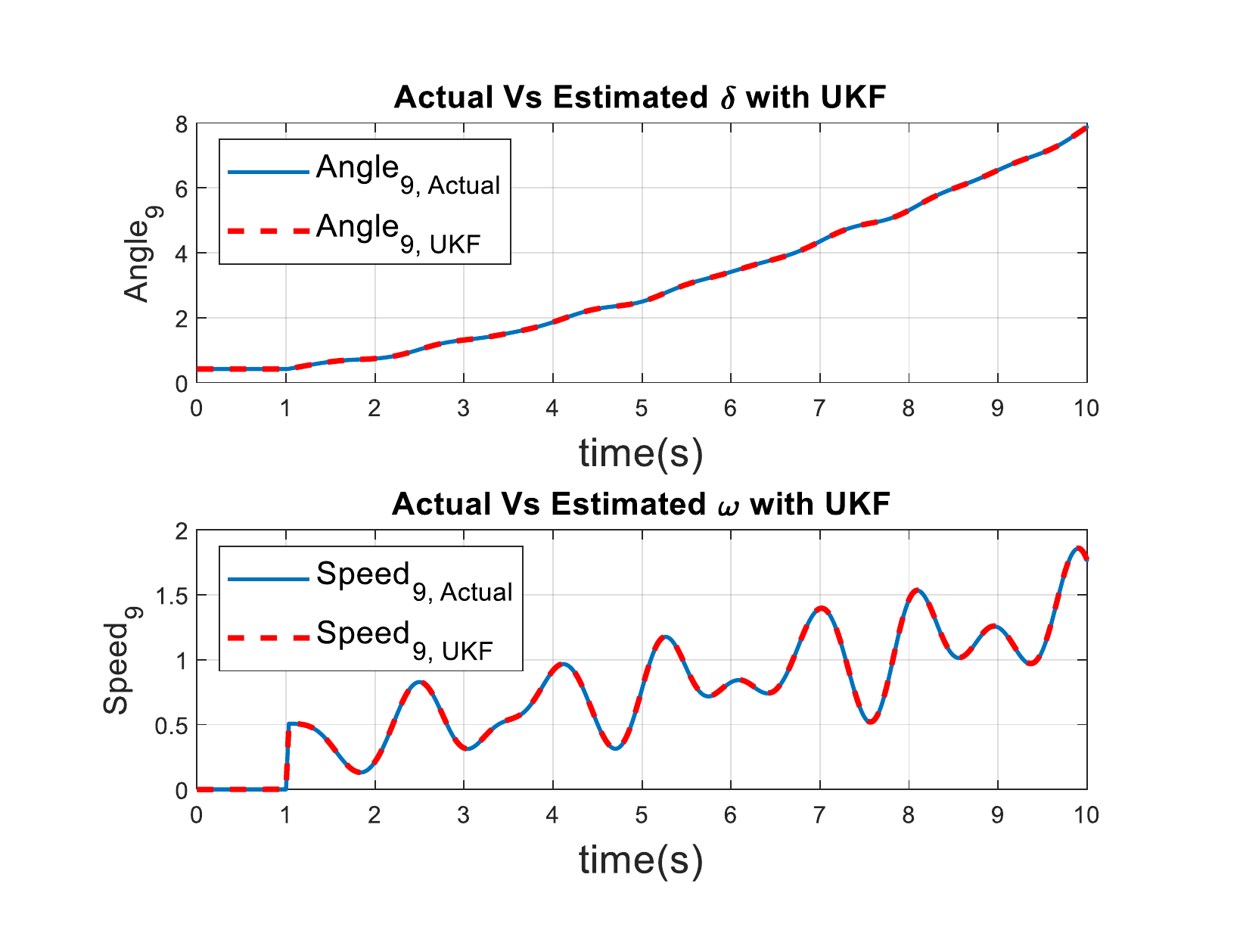}
    \caption{Actual vs estimated machine rotor angle and speed of generator 9 with UKF of New England 39 bus system.}
    \label{fig:39Generator_9_UKF}
\end{figure}

\section{Conclusion}\label{sec:conclusion}\label{sec:conclusion}
This paper estimated the power system dynamic states using extended and unscented Kalman filters. The case studies were performed on WECC $3$-machine $9$-bus system and New England $10$-machine 39-bus system. The results showed that both EKF and UKF can accurately estimate the power system dynamic states.

\bibliographystyle{IEEEtran}
\bibliography{References.bib}
\end{document}